\documentclass[aps,prd,longbibliography,twocolumn,floatfix,noeprint]{revtex4-2}

\usepackage{amsmath}
\usepackage{amssymb}
\usepackage{amsfonts}
\usepackage{graphicx}
\usepackage{dcolumn}
\usepackage{bm}
\usepackage{latexsym}
\usepackage{slashed}
\usepackage{xcolor}
\usepackage[colorlinks=true,linkcolor=blue,citecolor=blue]{hyperref}
\usepackage{array}

\newcommand{\Rco}{$\mathcal R_{\slashed E}$}
\definecolor{sand}{rgb}{0.76, 0.7, 0.5}

\begin{document}

\title{Effective theory for stochastic particle acceleration, with application to magnetized turbulence}

\author{Martin Lemoine} 
\affiliation{Astroparticule et Cosmologie, CNRS, Université Paris Cité,  F-75013 Paris, France}

\date{\today}

\begin{abstract}  
The physics of particle acceleration in turbulent plasmas is a topic of broad interest, which is making rapid progress thanks to dedicated, large-scale numerical experiments. The first part of this paper presents an effective theory of stochastic Fermi acceleration, which subsumes all forms of non-resonant acceleration in ideal electric fields and is applicable in generic settings. It combines an exact equation connecting the energization rate to the statistics of the velocity field with a statistical model of particle transport through the structures (i.e., the regions of strong velocity gradients). In a second part, this formalism is applied to magnetohydrodynamic turbulence to obtain a comprehensive assessment of the scale-by-scale contributions to the advection and diffusion coefficients. Acceleration peaks on scales where particles can be trapped inside structures for an eddy turn-around time, or in intense structures associated with sharp bends of the magnetic field lines in large-amplitude turbulence (as reported earlier). These spatially inhomogeneous, rapid acceleration regimes pave the way for a rich phenomenology. We discuss the scalings obtained, their interpretation and show that the findings compare satisfactorily with existing numerical results.
\end{abstract}

\maketitle

\section{Introduction}
Stochastic particle acceleration denotes the process by which charged particles gain energy through repeated random interactions with ideal electric fields~\cite{1949PhRv...75.1169F,1954ApJ...119....1F}. {\it Per se}, this definition opens up a broad field of applications, especially in astrophysics where disordered magnetized plasmas and supra-thermal populations are common occurences. Particles scattering on plasmoids in multi-island magnetic reconnection, e.g.~\cite{2006Natur.443..553D,2012PhRvL.108x1102K,PhysRevLett.110.151101,2021PhRvL.126m5101A,2022NatRP...4..263J}, particles diffusing in large-scale compressive motions of solar to interstellar and extragalactic plasmas, e.g.~\cite{96Larosa,2017ApJ...850L..28L,1983ICRC....9..313B,1988SvAL...14..255P,2011PhRvL.107i1101M,2007MNRAS.378..245B,2023ApJ...947...89L}, in sheared velocity flows, e.g.~\cite{1981SvAL....7..352B,2006ApJ...652.1044R,2018ApJ...855...31W,2019Galax...7...78R}, or even extreme explosive phenomena, e.g.~\cite{2016PhRvD..94b3005A,2017ApJ...846L..28X,2021PhRvD.104j3005Z}, provide but a few settings and realizations of the stochastic Fermi process.

While the present work aims to develop tools for modeling stochastic acceleration in generic environments, it will focus primarily on its application to magnetohydrodynamic (MHD) turbulence, which provides one of its most natural playgrounds. In this context, a large body of literature has focused on energization through resonant wave-particle interactions, starting with~\cite{66Kennel,67Hall,68Lerche,1971ApSS..12..302K}, and the corresponding transport coefficients are aptly derived within the framework of quasilinear theory, e.g.~\cite{2002cra..book.....S}. However, there is no consensus on which mechanism (resonant versus nonresonant) prevails in general, even less so depending on the physical conditions at hand, characterized by the plasma beta, the turbulence amplitude and the characteristic eddy velocity. 

This dualistic view is reminiscent of a long-standing similar debate in the turbulence community regarding the role of waves versus structures~\cite{2015RSPTA.37340154M,2019PhRvX...9c1037G,2023PhPl...30d0502V}.  With regard to particle acceleration, the distinction between these pictures is often nebulous. Transit-time damping acceleration, for instance, can be pictured both as a particle moving in resonance with the phase velocity of a small-amplitude magnetosonic wave, or as the deflection of that particle in the rest frame of a mirror perturbation of finite amplitude~\cite{1981A&A....97..259A,98Schlick}. If the wavelength of a mode exceeds the scattering mean free path of the particle, or if the amplitude of the mode becomes large enough to trap the particle, then particle acceleration becomes truly non-resonant, even in wave-like modes. From an observational point of view, a wavepacket of limited extent would be regarded as wave-like if its polarization matches that of e.g., Alfvén waves~\cite{2013LRSP...10....2B}, yet this limited extent would prevent initiating a resonant response~\cite{2012ApJ...744...75R}, and hence it would behave instead as a structure from the point of view of particle acceleration. 

While numerical (kinetic or fluid) simulations offer powerful tools to model turbulence and study the physics behind particle acceleration, a clear picture has yet to emerge. Regarding the nature of the fluctuations,  recent MHD simulations have provided evidence for dispersion-broadened dispersion relations of the MHD eigenmodes~\cite{2023arXiv231003806Y}. However, these modes fail to explain all observed turbulent energy, and other studies have provided different views on this issue, e.g.~\cite{2010PhPl...17j2304P,2010JGRA..115.4101N,2017PhRvE..96b3201P,2019PhRvX...9c1037G}. A lingering question is whether wave modes are randomly phased, as assumed in wave-particle resonant interactions, or whether coherence builds up. Various simulations tracking the history of particles in MHD flows suggest that coherent structures play an important role in particle energization~\cite{2004ApJ...617..667D,06Arzner,2013ApJ...777..128L,14Lynn,2014ApJ...783..143D,2017PhRvL.119d5101I,2018JPlPh..84f7201P,2019MNRAS.485..163K,2020ApJ...894..136T,2021MNRAS.506.1128S,2022ApJ...928...25P,2023ApJ...959...28P,2024A&A...686A.116P}. The hybrid and MHD simulations of Refs.~\cite{2020ApJ...894..136T,2022ApJ...928...25P}, in particular, have clearly shown that particle acceleration appears faster and more intense in specific regions of space. Based on the comparison between MHD simulations with and without the inclusion of nonlinear terms, with and without random phases, the recent study~\cite{2023ApJ...959...28P} concludes as well that structures regulate most of the particle acceleration. Fully kinetic (particle-in-cell, PIC) simulations conducted in the relativistic regime have provided first-principle views on the injection and acceleration process in large-amplitude turbulence~\cite{17Zhdankin,2018ApJ...867L..18Z,18Comisso,2019ApJ...886..122C,2019PhRvL.122e5101Z,2020ApJ...893L...7W,2021ApJ...922..172Z,2022ApJ...924L..19V,2022PhRvD.106b3028B,2025arXiv250203042W}. An interesting finding of these simulations is that, in the relativistic regime $v_{\rm A}\simeq c$, the diffusion coefficient scales as the squared Alfvénic 4-velocity $u_{\rm A}^2\equiv v_{\rm A}^2/(1-v_{\rm A}^2/c^2)$, rather than $(v_{\rm A}/c)^2$. By itself, this points toward a generalized Fermi process involving relativistically moving velocity structures~\cite{2019PhRvD..99h3006L,2021PhRvD.104f3020L}, rather than wave-particle interactions. On the basis of direct measurements in a MHD simulation of turbulence without guide field, Refs.~\cite{2021PhRvD.104f3020L,2022PhRvD.106b3028B,2022PhRvL.129u5101L} have further shown that acceleration is dominated by the interactions with intense structures characterized by sharp bends of the magnetic field lines and compression modes perpendicular to the field. These findings do not prejudge, however, the physics of acceleration in the sub-relativistic regime or at low Mach numbers. 

Most theoretical studies discussing non-resonant stochastic turbulent acceleration have focused on the role of large-scale compressive motions of the plasma. Key contributions include Refs.~\cite{1983ICRC....9..313B,1988SvAL...14..255P}, which have derived the first expressions for the diffusion coefficients of particles exploring large-scale compressive modes, relying on a perturbative expansion of the kinetic transport equation. This approach has been further developed in Ref.~\cite{2003ApJ...595..195W} and applied to other contexts such as turbulent reconnection~\cite{2015ApJ...801..112L}. In Refs.~\cite{2003ApJ...599.1426C,2004ApJ...603...23C,2006ApJ...638..811C}, the authors assessed the contributions of compressive modes across various scales, utilizing a microscopic formulation to describe the random walk of a particle, incorporating possible anisotropy and finite lifetime effects. This list of references and methods is certainly not exhaustive, especially when considering stochastic acceleration in a broader context, as touched upon at the beginning of this Introduction.

In this broad context, the present paper aims to advance our understanding in two ways. First, we propose a general, effective theory of stochastic acceleration that can be applied to a generic setting, including MHD turbulence but also all forms of structured or random velocity fields seeded by various types of plasma instabilities. This model, which forms the focus of Sec.~\ref{sec:effth}, is a non-perturbative microscopic formulation that holds equally well in relativistic as in non-relativistic settings. It builds up and generalizes our previous work~\cite{2019PhRvD..99h3006L,2021PhRvD.104f3020L} to provide an exact equation, Eq.~(\ref{eq:dyn_p}), which connects the rate of energization to the presence of velocity structures and their nature. It complements this equation with a statistical model describing the properties of the structures (e.g., their coherence time and oscillating frequency) and the kinematics of the interaction (e.g., how particles cross the structure), borrowing on the above previous studies and others to provide a general tool for calculating the transport coefficients. In Sec.~\ref{sec:MHDturb}, this formalism is applied to the context of MHD turbulence, and it is used to derive predictions for the transport coefficients and the contributions of different modes, scale by scale. These findings are summarized and discussed in detail in Sec.~\ref{sec:disc}, which also compares the results to numerical simulations.

Before introducing this effective theory, we recall a few fundamental notions on the meaning and the role of the transport coefficients. In generic applications of stochastic acceleration, the objective is often to solve a transport equation written in a Fokker-Planck form, characterized by the advection coefficient $A_p\equiv \langle\Delta p\rangle/\Delta t$ and the diffusion coefficient $D_{pp}\equiv \langle\Delta p^2\rangle/\Delta t$. This Fokker-Planck form is an approximation that is valid in the limit of short coherence time of the random force and small excursions in momentum per interaction. It is therefore not guaranteed to apply, at least on short or intermediate timescales before central limit convergence comes into play. Indeed, some numerical simulations have reported deviations from this Fokker-Planck behavior, e.g.~\cite{2020ApJ...894..136T,2021PhRvD.104f3020L,2022PhRvL.129u5101L}. However, for clarity, we focus here on determining $A_p$ and $D_{pp}$ to extract the dominant scale contributions, noting that the probability distribution function (pdf) of the momentum jumps can be deduced from the obtained formulas. We revisit the interpretation of these transport coefficients and the correct implementation of the transport equation in Sec.~\ref{sec:transport}.

\section{Effective theory for Fermi acceleration}\label{sec:effth}
The effective model of generalized Fermi acceleration that we present here builds upon Refs.~\cite{2019PhRvD..99h3006L,2021PhRvD.104f3020L}, which it generalizes in several ways, as described thereafter. The guiding principle of this method is to track the particle momentum in a sequence of reference frames \Rco\ in which the electric field vanishes at all points along the particle trajectory. In line with the original Fermi scenario, we assume throughout that the electric field is ``ideal''. This means that the two electromagnetic Lorentz invariants satisfy $\boldsymbol{E}^2-\boldsymbol{B}^2<0$ and $\boldsymbol{E\cdot B}=0$, as is the case for ideal MHD. In this case, \Rco\ is guaranteed to exist. The present formalism could also be generalized to accommodate a  non-ideal (parallel to $\boldsymbol{B}$) electric field component by adding its contribution to Eq.~(\ref{eq:dyn_0}) and subsequent equations.

The change of reference frame that we introduce enables the derivation of an exact equation [Eq.~(\ref{eq:dyn_p})] linking the rate of energy gain to the statistics of the gradients of the velocity field $\boldsymbol{v_E}$ (velocity of \Rco\ in the laboratory frame). This formalism provides a direct generalization of the Fermi process, because exploring gradients in the velocity field ensures that particles interact with electric fields that cannot be erased by a global Lorentz transformation, and such is the true nature of Fermi acceleration. Moreover, it offers a fully covariant description of acceleration, which can therefore be applied to relativistic turbulence. This formalism is detailed in the following Sec.~\ref{sec:dyneq}.

The dynamical Eq.~(\ref{eq:dyn_p}) formally takes the form of a stochastic equation driven by the random gradients of $\boldsymbol{v_E}$, which parallels the equation written in the laboratory frame describing the interactions with the random electric field $\boldsymbol{E}$, i.e.  $\dot p = q\, \boldsymbol{\hat p}\cdot\boldsymbol{E}$ ($\boldsymbol{\hat p}\equiv \boldsymbol{p}/|\boldsymbol{p}\vert$, with $\boldsymbol{p}$ the particle momentum in the laboratory frame, and $q$ is the particle charge). To compute the transport coefficients, one integrates such equations forward in time to derive moments of the form $\langle\Delta p^m\rangle$ ($m=1,2$). These terms involve the $m$-point correlation function of the stochastic force term (i.e., $\boldsymbol{E}$ in the laboratory frame, or gradients of $\boldsymbol{v_E}$ in our \Rco\ formulation) \emph{along the particle trajectory}. Consequently, the computation requires modeling the statistical properties of particle transport in order to connect these $m$-point functions to the statistical properties of the random force fields. For example, in standard quasilinear calculations, one proceeds to lowest order of perturbation in $\delta E$ and describes the transport as unperturbed gyration around the mean magnetic field. The second part (Sec.~\ref{sec:transport}) of the discussion addresses this issue, borrowing from previous work to eventually provide a tool for extracting the transport coefficients from the dynamical equation.

\subsection{Dynamical equations}\label{sec:dyneq}
We thus track the evolution of the momentum in a frame \Rco\ in which the electric field $\boldsymbol{E}=\boldsymbol{0}$, and we denote by $\boldsymbol{v_E}$ the velocity of this frame relative to the laboratory frame. At every point in space, there exists an infinite number of \Rco\ frames, because in the ideal MHD approximation, $\boldsymbol{E}$ vanishes in any frame whose velocity perpendicular to the magnetic field line is equal to $\boldsymbol{v_B}=c\,\boldsymbol{E}\times\boldsymbol{B}/B^2$, the velocity of magnetic field lines ($\boldsymbol{B}$ denoting the total magnetic field). This leaves the velocity along $\boldsymbol{B}$ unspecified. The frame of rest of the magnetic field lines thus provides one such frame, as does the plasma rest frame, which drifts perpendicularly to $\boldsymbol{B}$ at velocity $\boldsymbol{v_B}$, as well as along $\boldsymbol{B}$. The full plasma velocity is hereafter written $\boldsymbol{v_{\rm p}}$. Which frame to adopt is a matter of convenience, which will be specified later on. To anticipate slightly, it appears natural to choose \Rco\ with $\boldsymbol{v_E} = \boldsymbol{v_B}$ when describing the acceleration of a particle tied to a magnetic field line, while $\boldsymbol{v_E} = \boldsymbol{v_{\rm p}}$ becomes more natural when the transport is modeled by diffusion on small-scale perturbations carried by the flow. For now, $\boldsymbol{v_E}$ is left unspecified. Quantities defined in \Rco\ are marked with primed symbols, e.g. the particle momentum ${p'}$. We also introduce the four-velocity $u_E^\mu$ associated with $\boldsymbol{v_E}$, $\boldsymbol{u_E}=\gamma_E \boldsymbol{v_E}$ with $\gamma_E\equiv 1/(1-v_E^2/c^2)^{1/2}$ and ${u_E}^\mu\,{u_E}_\mu\,=\,-c^2$.

The evolution of the particle momentum in the (non-inertial) frame \Rco\ can be derived using the techniques introduced in Refs.~\cite{1985ApJ...296..319W,1989ApJ...340.1112W,2019PhRvD..99h3006L,2021PhRvD.104f3020L}. The first step is to establish a tetrad $\{{e_a}^\mu\}$ that connects the two frames at any point in space. In particular, any four-vector in the laboratory frame, e.g. the particle four-momentum $p^\mu$, can be written in terms of its equivalent in the comoving frame through the correspondence $p^\mu\,=\,{e_a}^\mu {p'}^a$. Following earlier notations, greek indices ($\alpha$, $\beta$, etc.) denote space-time coordinates in the laboratory frame, while latin indices represent coordinates in the locally inertial frame defined by the tetrad. Indices $a,\,b,\ldots$ span $0,\ldots,3$, and indices $i,\,j,\ldots$ cover the spatial part $1,\ldots,3$. Unlike Ref.~\cite{2021PhRvD.104f3020L}, we leave this tetrad mostly unspecified, but we retain ${u_E}^\mu$ as the timelike direction, meaning ${e_0}^\mu\equiv {u_E}^\mu/c$. We further choose the set as orthonormal, i.e. $\eta_{\mu\nu} {e_a}^\mu {e_b}^\nu = \eta_{ab}$, with $\eta_{ab}$ the Minkowski metric (signature $-,+,+,+$). The spacelike components ${e_i}^\mu$ ($i\in{1,2,3}$) that define the spacelike directions at each point are not specified for now. Following standard conventions, factors of $c$ are included in the time component of four-vectors, e.g. ${u_E}^\mu\,=\,(\gamma_E c,\,\gamma_E\boldsymbol{v_E})$ and $\partial_\mu\,=\,(c^{-1}\partial_t,\,\boldsymbol{\nabla})$.

The general equation of evolution in \Rco\ takes the form ($m$ denoting the mass of the particle)
\begin{equation}
\frac{{\rm d} {p'}^a}{{\rm d}\tau}\,=\,\frac{q}{m}\,
{F^a}\!_b\, {p'}^b\,+\,{{e_b}^\beta}{{e_c}^\gamma}{{e^a}_{\beta,\gamma}}
\,\frac{{p'}^b\,{p'}^c}{m}\,.
\label{eq:dyn}
\end{equation}
The Lorentz force acting in the comoving frame is encoded in the Maxwell field strength tensor ${F^a}\!_b$. The second term on the right hand side describe the inertial forces associated with the non-inertial nature of \Rco. In particular, setting $a=0$ to describe the evolution of the energy, the Lorentz force vanishes (by construction of \Rco) so that the energy gains and losses result from the inertial forces characterized by the spatio-temporal gradients of the velocity of \Rco\, since ${e^0}_{\beta,\gamma}=-{u_E}_{\beta,\gamma}/c$.

We now write down the equation of evolution of the energy $\epsilon'\equiv{p'}^0\,c$ and of a spatial component ${p'}^i$. We also adopt $t'$ as the local time coordinate instead of proper time $\tau$  used in Eq.~(\ref{eq:dyn}), noting that ${\rm d}\tau = {\rm d}t'\,mc^2/\epsilon'$. First, for $\epsilon'$, with implicit summation over the spatial indices:
\begin{align}
    \frac{{\rm d}\epsilon'}{{\rm d}t'}\,=\,&-\frac{1}{\epsilon'}{u_E}_{\beta,\gamma}\biggl\{
    {u_E}^\gamma{e_i}^\beta\,\epsilon'{p'}^i+ {e_i}^\beta {e_j}^\gamma\,{{p'}^i}{{p'}^j} c^2\biggr\}\,.
    \label{eq:dyn_0}
\end{align}
Each term has a specific interpretation in terms of the kinematic decomposition of the spatio-temporal derivatives of ${u_E}^\mu$, i.e. the acceleration, the shear, the vorticity and the compression. The first term, in particular, is directly linked to the acceleration of the \Rco\ frame, defined by ${a_E}_\beta \equiv {u_E}^\gamma\,{u_E}_{\beta,\gamma}$. The corresponding term represents the inertial force associated with this acceleration projected onto the space-like directions ${e_i}^\beta$. Hereafter, this term is written $a_i\equiv c^{-1}\,{a_E}_\beta {e_i}^\beta$, including an additional factor of $c^{-1}$ for dimensional consistency. 

The quantity ${u_E}_{\beta,\gamma}{e_i}^\gamma$ stands for the derivative of ${u_E}_\beta$ taken along the direction set by ${e_i}^\gamma$, so that $\sigma_{ij}\equiv {u_E}_{\beta,\gamma}{e_i}^\beta {e_j}^\gamma$ represents the shear of the four-velocity in the spatial plane spanned by ${e_i}^\beta$ and ${e_j}^\gamma$~\footnote{In a kinematical decomposition, the shear $\sigma_{ij}$ is formally defined as the traceless symmetric part of ${u_E}_{\beta,\gamma}{e_i}^\beta {e_j}^\gamma$, and the vorticity as the antisymmetric part. However, the vorticity cancels out in Eq.~(\ref{eq:dyn_0}). For simplicity, and to make contact with more common definitions of the shear in models of particle acceleration, we retain a specific ordering for the indices: as defined in the text, $\sigma_{ij}$ here corresponds to the derivative of the velocity component along ${e_i}^\beta$ in the direction set by ${e_j}^\gamma$.}. The diagonal part of the shear tensor, namely ${u_E}_{\beta,\gamma}{e_i}^\gamma {e_i}^\beta$ (no sum over $i$ intended) characterizes compression in the direction set by ${e_i}^\beta$. Hence, the second term inside the curly brackets includes both compressive terms in the various spatial directions (diagonal part) and shear contributions ($i\neq j$). The compressive terms are hereafter written $\Theta_i \equiv {u_E}_{\beta,\gamma}{e_i}^\beta {e_i}^\gamma$ (no summation over $i$ intended).

For clarity, all sums over spatial indices are now made explicit. Introducing the three velocity components ${v'}_i\equiv {p'}_i c^2/\epsilon'$, the above expressions become
\begin{align}
\frac{{\rm d}\epsilon'}{{\rm d}t'}\,=\,& - \sum_{i=1}^{3}\left\{{p'}_ic\,a_i + {p'}_i {v'}_i\, \Theta_i + \sum_{j\neq i}{p'}_i {v'}_j\,\sigma_{ij}\right\}\,, 
\label{eq:dyn_e}
\end{align}
or, in terms of momentum $p'$ and the angular variables ${\mu'}_i\equiv {p'}_i/p'$,
\begin{align}
\frac{{\rm d}p'}{{\rm d}t'}\,=\,& - \sum_{i}\left\{\frac{1}{c}\epsilon' {\mu'}_i a_i + {p'}{{\mu'}_i}^2 \Theta_i + \sum_{j\neq i}p' {\mu'}_i  {\mu'}_j\sigma_{ij}\right\}\,, 
\label{eq:dyn_p}
\end{align}
which offers a simple and comprehensive account of all possible non-resonant energization channels in the present framework of ideal MHD. Standard mechanisms such as betatron acceleration, curvature and gradient drifts, transit-time damping and magnetic pumping all emerge as limiting cases of this equation, as will become apparent over the course of the discussion in Sec.~\ref{sec:MHDturb}. This equation is exact at every point in space (as long as \Rco\ exists), and it does not make any assumption as to the properties of transport of the particle. 

It also generalizes the formalism introduced in Refs.~\cite{2019PhRvD..99h3006L,2021PhRvD.104f3020L} as follows. While Ref.~\cite{2019PhRvD..99h3006L} directly integrated Eq.~(\ref{eq:dyn}) assuming isotropic diffusion of particles, Ref.~\cite{2021PhRvD.104f3020L} adopted the specific choice $\boldsymbol{v_E}=\boldsymbol{v_B}$ to describe the acceleration of particles in gyration around a mean magnetic field line and oriented ${e_1}^\mu$ along the magnetic field direction, i.e. ${e_1}^\beta \equiv b^\beta \equiv B^\beta/\sqrt{B^\alpha B_\alpha}$, with $b^\beta$ denoting the magnetic field line direction four-vector. The term $\Theta_1$ then describes the compression of ${u_E}^\mu$ measured along the magnetic field. Energization (or de-energization) occurs when the particle crosses a bend of the magnetic field with $\boldsymbol{u_E}$ in the plane of curvature, which is equivalent to curvature drift acceleration. Assuming further that gyration takes place on short timescales compared to the scales of variation of the velocity field, one can approximate terms involving $p_2'$ and $p_3'$ by their gyroaverages, giving $\langle {p'}_2{p'}_3\rangle \simeq 0$, $\langle {{p'}_2}^2\rangle\simeq\langle {{p'}_3}^2\rangle\simeq {{p_\perp'}^2}/2$, which removes the terms proportional to the shear components $\sigma_{ij}$ and transforms the combination of 
${{p'}_2}^2\Theta_2$ and ${{p'}_3}^2\Theta_3$ into a compression transverse to the field line. This terms incorporates both betatron and gradient-drift acceleration~\cite{2021PhRvD.104f3020L}. 

Nonetheless, the terms that appear in Eqs.~(\ref{eq:dyn_e}) and (\ref{eq:dyn_p}) do not make any reference to the conservation of an adiabatic invariant, and the present approach is not a guiding center description. Consider for instance the limit in which the pitch angle cosine ${\mu'}_1$ is randomized by scattering events while the particle crosses a bend of the magnetic field line. Scattering violates the adiabatic invariant, yet Eq.~(\ref{eq:dyn_p}) shows that particles remain energized by the dynamic curvature, according to $\dot{p'}\simeq -p'{{\mu'}_1}^2\Theta_1$. Similarly, in the sub-relativistic limit ($\gamma_E\simeq 1$), the assumption of fast scattering $\nu_{\rm s}\rightarrow+\infty$ ($\nu_{\rm s}$ is the scattering frequency) implies on average ${\mu'}_i\rightarrow 0$ but 
${{\mu'}_i}^2\rightarrow 1/3$, thereby operating the transition to the fluid limit,  transforming Eq.~(\ref{eq:dyn_p}) into \footnote{As ${e^i}_\beta {e_i}^\gamma\,=\,{\delta_\beta}^\gamma - {u_E}_\beta {u_E}^\gamma$, one can write $\sum_i \Theta_i = \partial_\alpha {u_E}^\alpha$, which reduces to $\boldsymbol{\nabla}\cdot\boldsymbol{v_E}$ in the sub-relativistic limit $v_E\ll c$.}
\begin{equation}
    \frac{{\rm d}{p'}}{{\rm d}t'}\,=\,-\frac{1}{3}\,{p'}\,\boldsymbol{\nabla}\cdot\boldsymbol{v_E}\,.
    \label{eq:dyn-fluid}
\end{equation}
This equation forms the basis of Refs.~\cite{2003ApJ...595..195W,2003ApJ...599.1426C,2004ApJ...603...23C,2006ApJ...638..811C} for their studies of particle acceleration in turbulent compressive modes. The present formulation therefore generalizes these studies as well.

Considering now the evolution of a specific momentum component ${p'}^i$, Eq.~(\ref{eq:dyn}) can be partially expanded and rewritten as
\begin{align}
\frac{{\rm d}{p'}^i}{{\rm d}t'}&\,=\, {e^i}_{\beta,\gamma}\biggl\{{u_E}^\beta{u_E}^\gamma \,\frac{\epsilon'}{c^2}  + \sum_{j}{u_E}^\beta {e_j}^\gamma \, {p'}_j \nonumber\\
 & \quad\quad\quad+{e_c}^\gamma {p'}^c\sum_{k\neq i} {v'}_k{e_k}^\beta \biggr\}\nonumber\\
 & \quad+ \mathrm{Lorentz}\,,
    \label{eq:dyn_u1_0}
\end{align}
which is organized as follows. First, only the first two terms (four when expanding the  sum over $j$)  inside the curly brackets contribute to energization. Noting that 
${e^i}_{\beta,\gamma}\,{u_E}^\beta = -{e^i}_\beta\,{{u_E}^\beta}_{,\gamma}$ because ${e^i}_\beta {u_E}^\beta=0$, one identifies the first of these four terms with the acceleration $a_i$, the term $j=i$ with $\Theta_i$, and the remaining two with $\sigma_{ij}$ ($j\neq i$). The remaining terms inside the curly brackets describe the mixing of momentum components through a possible rotation of the spatial axes of the reference system, because ${e_c}^\gamma\,{p'}^c= p^\gamma$, and $p^\gamma {e^i}_{\beta,\gamma}/mc$ represents the derivative of ${e^i}_\beta$ taken along the particle trajectory. Finally, the term  ``Lorentz'' includes any possible gyration associated with the presence of a magnetic field in the comoving frame. This term disappears if ${e^i}_\beta$ is identified with the magnetic field line direction. In summary,
\begin{align}
\frac{{\rm d}{p'}^i}{{\rm d}t'}\,=\,&-\frac{1}{c}\epsilon'a_i - {p'}_i\Theta_i - \sum_{j\neq i}{p'}_j\sigma_{ij} \nonumber\\
& + p^\gamma {e^i}_{\beta,\gamma}\sum_{k\neq i}{v'}_k{e_k}^\beta + \mathrm{Lorentz}\,.
\label{eq:dyn_u1_1}
\end{align}
Alternatively, from Eqs.~(\ref{eq:dyn_p}) and (\ref{eq:dyn_u1_1}), we can extract the evolution of ${\mu'}_i$ (no sum over $i$ intended) 
\begin{align}
\frac{{\rm d}{\mu'}_i}{{\rm d}t'}&\,=\,-\frac{c}{v'}(1-{{\mu'}_i}^2)\,a_i -
{\mu'}_i(1-{{\mu'}_i}^2)\,\Theta_i \nonumber\\
& -\sum_{j\neq i}{\mu'}_j(1-{{\mu'}_i}^2)\,\sigma_{ij} \nonumber\\
&+\sum_{j\neq i}{{\mu'}_i}^2{\mu'}_j\,\sigma_{ji} + \sum_{j\neq i,k\neq i,k\neq j}{\mu'}_i{\mu'}_j{\mu'}_k\,\sigma_{jk} \nonumber\\
&+\sum_{j\neq i}\frac{c}{v'}{\mu'}_i{\mu'}_j\,a_j   + \sum_{j\neq i}{\mu'}_i{{\mu'}_j}^2\,\Theta_j \nonumber\\
& + \sum_{j\neq i}\frac{c^2}{\epsilon'}p^\gamma {e^i}_{\beta,\gamma}{\mu'}_j{e_j}^\beta  + \mathrm{Lorentz} \,,
\label{eq:dyn_mu1_0}
\end{align}
The last two terms do not directly scale with ${u_E}^\mu$. They therefore dominate angular scattering in the sub-relativistic ($u_E\ll c$) limit, while the other terms indicate how angular variables change under the influence of the energizing forces. Additionally, the penultimate term includes a contribution $\propto {e^i}_{\beta,\gamma}\sum_{j,k\neq i} {{p'}_j}{{p'}_k}{e_j}^\beta {e_k}^\gamma $ (when expanding $p^\gamma = p^a {e_a}^\gamma$). When ${e^i}_\beta$ is oriented along the magnetic field line direction and a gyrophase average is carried out, this term becomes the mirror force~\cite{2021PhRvD.104f3020L}.

\subsection{Transport and interaction with structures}
The system of Eqs.~(\ref{eq:dyn_p}) and (\ref{eq:dyn_mu1_0}) or (\ref{eq:dyn_u1_1}), as well as similar equations for the other angular variables, forms an exact random dynamical system describing the co-evolution of energy and angular variables in the comoving frame, whose resolution gives access to the statistics of momentum jumps such as the advection and diffusion coefficients, respectively $A_p= \langle\Delta p\rangle/\Delta t$ and  $D_{pp}= \langle\Delta p^2\rangle/\Delta t$. Since the random force terms in these equations are expressed at the location of the particle, one must model the transport of the particles and specify the statistical properties of the structures that they traverse.

In Sec.~\ref{sec:MHDturb}, we examine the contribution of structures scale by scale. To do so, we adopt a coarse-grained view of the turbulence and define structures (velocity gradients) on scale $l$ as ${\Theta_i}_l\equiv G_l\star{\Theta_i}$ (and similarly for $a_i$, $\sigma_{ij}$), where $G_l$ represents a convolution kernel filtering out scales $<l$. The reference frame \Rco\ is then defined relative to the coarse-grained velocity flow and coarse-grained electromagnetic structure on scale $l$ by identifying $\boldsymbol{v_E}$ with $c \boldsymbol{E_l}\times\boldsymbol{B_l}/{B_l}^2$ plus a potential longitudinal component along $\boldsymbol{B_l}$~\cite{2021PhRvD.104f3020L}. One direct consequence is that the small scales $l'<l$ that have been filtered out can contribute to spatial transport, because their (random) velocities do not vanish in the \Rco\ frame defined on scale $l$. Such random motion portrays turbulent transport, whose importance relative to particle acceleration has notably been stressed in Ref.~\cite{2003ApJ...595..195W}. The corresponding diffusion coefficient on scales $l'$ takes the approximate form $\sim l'\delta v_{l'}$, where $\delta v_{\rm l'}$ represents the characteristic (three-)velocity of eddies on scale $l'$. The total contribution of all modes on scales $l'<l$ is thus dominated by the largest scales $\sim l$, with effective diffusion coefficient $\sim l\delta v_l$, up to a numerical coefficient $\lesssim 1$.

To characterize the random forces along the trajectory of the particle, we adopt a Lagrangian point of view and borrow the notion of dynamical decorrelation of the random force that has been introduced to describe the scattering of particles in solar wind turbulence~\cite{1994ApJ...420..294B}, and later used in the context of stochastic particle acceleration, 
e.g.~\cite{2003ApJ...599.1426C,2004ApJ...603...23C,2006ApJ...638..811C,2018JPlPh..84f7201P}. 
Decorrelation occurs as the particle exits the structure, or because the structure itself has a finite correlation time. The decorrelation rate (inverse of decorrelation time) is hereafter written $\nu_{\rm f}$ and approximated by the maximum of $\nu_{\rm esc}$ (the rate at which particles escape from the structure) and $\nu_{\rm f0}$ (the proper rate of decorrelation of the structure). To avoid excessive details at this stage, these quantities are left unspecified for now, and detailed when relevant. We also consider that the structure may alternate its polarity in time or space. This choice is motivated by the observation that the turbulent velocity field itself has bounded variations, therefore the gradients that characterize the structures are bound to change sign in space over a distance comparable to (or smaller than) the structure coherence length. Additionally, the inclusion of an oscillatory substructure allows us to use the present formalism to treat the case of wavepackets of finite amplitude. This leads us to describe the structure as a damped oscillator, with decorrelation frequency $\nu_{\rm f}$ and oscillating frequency $\omega_{\rm f}$. Denoting the random force term (${a_i}_l$, ${\Theta_i}_l$, ${\sigma_{ij}}_l$) by the generic notation $\Gamma_l$, we thus adopt the following statistics
\begin{align}
    \langle\Gamma_l(t'_1)\rangle &\,=\,0\,,\nonumber\\
    \langle\Gamma_l(t'_1)\Gamma_l(t'_2)\rangle & \,=\, \langle \Gamma_l^2\rangle\, e^{-\nu_{\rm f}\vert t'_1-t'_2\vert}\cos\left[\omega_{\rm f}(t'_1-t'_2)\right]\,,
    \label{eq:stat-gamma}
\end{align}
and we neglect cross-correlations between gradients of different types, but note that this is not a trivial assumption. While seemingly reasonable in the context of strong turbulence, it may be violated in certain settings where acceleration, compression and shear are correlated because they share the same source. The above formulation also assumes that the second moment of $\Gamma_l$ is well defined, which is a necessary condition for the diffusion coefficient to be properly defined. Deviations from this case will be discussed in Sec.~\ref{sec:transport}.

The apparent oscillatory part may result from proper oscillations at frequency $\omega_{\rm f0}$, as well as from the exploration of the spatial structure, which is  characterized here by a wavenumber $k_{\rm f0}$. The proper frequency $\omega_{\rm f0}$ needs not vanish in \Rco, since this frame corresponds to the instantaneous rest frame where the electric field vanishes, not to the frame comoving with the structure. 
It will suffice to approximate  $\omega_{\rm f}$ as the maximum of $\omega_{\rm f0}$ and $k_{\rm f0}v$ when the particle moves at velocity $v$ ballistically through the structure, or $k_{\rm f0}^2 \kappa$ if instead it diffuses with spatial diffusion coefficient $\kappa$, e.g.~\cite{1966PhFl....9.1773D}. Here as well, this will be made more specific later on. 

Spatial transport can take different forms depending on the ordering between the scale $l$ and  the scattering mean free path of the particles. To describe the evolution of the pitch angle variables ${\mu'}_i$, we first simplify Eq.~(\ref{eq:dyn_mu1_0}) by retaining all terms that contribute to energization, and grouping the last two in the form of a generic stochastic process describing an independent source of angular scattering, written $\xi_{\mu_i'}$. We denote by $\mu_{\{\xi\}_i}'$ the evolution of ${\mu'}_i$ under the influence of this scattering process $\xi_{\mu_i'}$ alone, i.e. neglecting the influence of the energizing terms. This process can be characterized by a time correlation function, describing scattering of particles on magnetic inhomogeneities at rate $\nu_{\rm s}$, as well as changes in the direction with respect to which ${\mu'}_i$ is defined (e.g. due to small-scale perturbations in the direction of the magnetic field lines)~\cite{1975RvGSP..13..547V,2003ApJ...590L..53M,2018JPlPh..84f7201P}, or small-scale turbulent diffusion as discussed above.

We thus define $\lambda_\parallel\equiv v/\nu_{\rm s}$ as the parallel mean free path to scattering of ${\mu'}_{\{\xi\}_i}$ (here, along the magnetic field), under the influence of $\xi_{\mu_i'}$ alone. In the simplest approximation possible, one models transport as isotropic with  
\begin{align}
    \langle {\mu'}_{\{\xi\}_i}\rangle &\,=\,0 \nonumber\\
    \langle {\mu'}_{\{\xi\}_i}(t_1)\,{\mu'}_{\{\xi\}_j}(t_2)\rangle &\,=\, \frac{1}{3}\,\delta_{ij}\, e^{-\nu_{\rm s}\vert t_1-t_2\vert}\,.
    \label{eq:stat-mu}
\end{align}
The above formulation applies if the magnetic fluctuation amplitude on the outer scale is at least comparable to the guide field, as otherwise, one needs to distinguish the (fast) perpendicular decorrelation rate from the parallel one. The generalization is straightforward and postponed to Sec.~\ref{sec:gfield}. The above formulation also neglects the influence of turbulent transport by smaller scale eddies, which will be properly implemented in Sec.~\ref{sec:large}.

The set of Eqs.~(\ref{eq:dyn_p}), (\ref{eq:dyn_mu1_0}), (\ref{eq:stat-gamma}) and (\ref{eq:stat-mu})  allows one to calculate the momentum transport coefficients in a generic setting. In a way, this formulation of stochastic Fermi-type acceleration parallels the calculations performed in the context of extended quasilinear theories, see e.g.~\cite{2020PhRvD.102b3003D} for a recent implementation, replacing the bath of random-phased linear waves with the statistics of velocity gradients. In the following, this formalism is used to calculate the advection and diffusion coefficients for Fermi-type stochastic acceleration in the context of MHD turbulence.

\section{Application to MHD turbulence}\label{sec:MHDturb}
For simplicity and ease of notation, we restrict the discussion to sub-relativistic turbulence and thus remove all prime symbols. In Sec.~\ref{sec:simnum}, we explain how the results that follow can be generalized to relativistic turbulence. The coherence scale of the turbulence is written $\ell_{\rm c}$ and the characteristic eddy velocity on this outer scale is $\delta v_{\rm c}$. The guide field is written $\boldsymbol{B_0}$, and the total magnetic field $\boldsymbol{B}$. Their amplitudes are denoted by $B_0$ and $B$, respectively; the average mean (total) magnetic field strength is written $\overline B$, while $\delta B$ corresponds to the root mean squared fluctuation amplitude on the outer scale.

\begin{figure}[h]
\includegraphics[width=0.95\columnwidth]{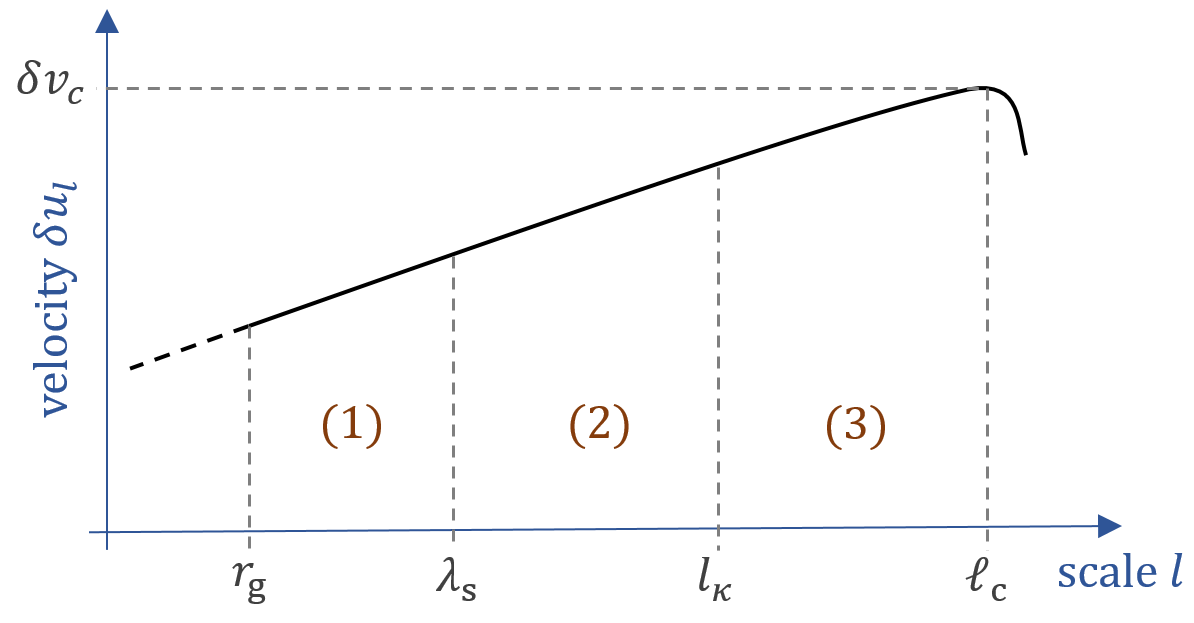}
\caption{\label{fig:f1} 
Illustration of the various length scales and the corresponding acceleration regimes. The vertical scale indicates the characteristic velocity fluctuation $\delta u_l$ on scale $l$ and a generic power-law scaling $\delta u_l = \delta v_{\rm c}(l/\ell_{\rm c})^{(q-1)/2}$, $q$ index of the power spectrum of velocity fluctuations (solid line). From smallest to largest: $r_{\rm g}$ is the particle gyroradius, $\lambda_{\rm s}$ is the particle mean free path to scattering, $l_\kappa$ is the transition scale to turbulent transport [Eq.~(\ref{eq:lkappa})], and $\ell_{\rm c}$ is the coherence scale of the turbulence (assuming here $l_\kappa< \ell_{\rm c}$; this ordering is discussed in Sec.~\ref{sec:disc}). For simplicity, this figure assumes a scale hierarchy characteristic of large-amplitude isotropic turbulence. The scattering mean free path $\lambda_{\rm s}=\lambda_\parallel=v/\nu_{\rm s}$. Depending on the scale $l$ of the mode under study, particle acceleration can be envisioned as follows: (1) the particle gyrates around the total magnetic field and is energized by curvature-drift, gradient-drift and betatron-type forces, as well as through the acceleration/deceleration of the field line, see Sec.~\ref{sec:interm}; (2) the particle crosses the perturbation diffusively and is subject to magnetic pumping in the compressive modes, as well as turbulent shear acceleration (and acceleration/deceleration of the plasma), see Sec.~\ref{sec:large}; (3) turbulent transport dominates and confines the particle in the mode for an eddy turn-around time, energization proceeding as in (2), see Sec.~\ref{sec:large}. For anisotropic or small-amplitude turbulence, see the text in Sec.~\ref{sec:aniso} and \ref{sec:gfield}. 
}
\end{figure}

To ease the reading, we provide a table (Table~\ref{tab:t1}) summarizing the main frequencies introduced in the text and their meaning, together with a figure (Fig.~\ref{fig:f1}) illustrating the different length scales and the corresponding regimes of acceleration.
\begin{table}[h]
\caption{\label{tab:t1}
Main frequencies introduced in the text, their associated meanings, and references to defining equations. In relation to the scattering frequency $\nu_{\rm s}$ of a particle, the text introduces the spatial diffusion coefficient (assuming isotropic scattering in large-amplitude turbulence) $\kappa_{\rm s}=v^2/\nu_{\rm s}$, $v$ denoting the particle velocity. Similarly, in relation to $\nu_{\rm t}$, the text defines a spatial diffusion coefficient for turbulent transport $\kappa_{\rm t}=\delta u_l^2/\nu_{\rm t}\simeq l\delta u_l$. The quantity $R_{\omega/\nu} = (1+\omega_{\rm f}^2/\nu_{\rm f}^2)$, see Eq.~(\ref{eq:def-Ron}).
}
\begin{ruledtabular}
\begin{tabular}{l  >{\raggedright}p{0.75\linewidth} r}
\colrule
$\nu_{\rm f}$ & decorrelation frequency of a random force $\Gamma_l$& Eq.~(\ref{eq:stat-gamma})\\ 
$\omega_{\rm f}$ & oscillating frequency of a random force $\Gamma_l$& Eq.~(\ref{eq:stat-gamma})\\ 
$\nu_{\rm s}$ & scattering frequency of a particle& Eq.~(\ref{eq:stat-mu})\\ 
$\nu_{\rm t}$ & $\mu$-decorrelation rate through turbulent transport & Eq.~(\ref{eq:stat-mu-turb})\\ 
\end{tabular}
\end{ruledtabular}
\end{table}

\subsection{Contribution from intermediate scales}\label{sec:interm}
On intermediate length scales $r_{\rm g}\lesssim l\lesssim \lambda_\parallel$, where $r_{\rm g}\equiv pc/ e\overline B$ denotes the gyroradius in the average total magnetic field, particles explore the structure by gyrating around the background magnetic field at parallel velocity $v_\parallel = v \mu$, with $\mu \equiv {\mu'}_1$ the pitch-angle cosine. In this limit, it becomes natural to orient ${e_1}^\mu$ along the magnetic field direction and set $\boldsymbol{v_E} = \boldsymbol{v_B}$, the velocity field of magnetic field lines. The variable ${\mu_{\{\xi\}_1}}$ can be regarded as approximately constant, while ${\mu_{\{\xi\}_2}}$ and ${\mu_{\{\xi\}_3}}$ are approximated by their gyrophase average, namely $\langle {\mu_{\{\xi\}_2}}\rangle = \langle {\mu_{\{\xi\}_3}}\rangle = \langle {\mu_{\{\xi\}_2}}{\mu_{\{\xi\}_3}}\rangle = 0$, $\langle {{\mu_{\{\xi\}_2}}}^2\rangle = \langle {\mu_{\{\xi\}_3}}^2\rangle = (1 - {{\mu_{\{\xi\}_1}}}^2)/2$. 

As recalled above, the forces acting on the particle can be written in this limit in terms of the longitudinal compression ${\Theta_\parallel}_l$ ($\equiv{\Theta_1}_l$) of the velocity field $\boldsymbol{{v_E}_l}$ ($\simeq \boldsymbol{{u_E}_l}$ in the sub-relativistic limit), the perpendicular compression ${\Theta_\perp}_l$ ($\equiv {\Theta_2}_l+{\Theta_3}_l$) and the longitudinal acceleration ${a_\parallel}_l$ ($\equiv {a_1}_l$). This description boils down -- but is not equivalent, see the discussion above -- to a stochastic guiding center approach in the turbulent field coarse-grained on scales $l$ larger than the particle gyroradius. In this description, ${\Theta_\parallel}_l$ embodies curvature drift acceleration, while the perpendicular gradient ${\Theta_\perp}_l$ encapsulates betatron and gradient drift acceleration~\cite{2021PhRvD.104f3020L}. The detailed expressions for these terms are 
\begin{align}
{\Theta_\parallel}_l&\,\simeq\, \boldsymbol{b_l}\cdot\left(\boldsymbol{b_l}\cdot\boldsymbol{\nabla}\right)\boldsymbol{{v_B}_l}\,\nonumber\\
{\Theta_\perp}_l&\,\simeq\,\boldsymbol{\nabla}\cdot\boldsymbol{{v_B}_l}-{\Theta_\parallel}_l\nonumber\\
{a_\parallel}_l &\,\simeq\,\frac{1}{c}\boldsymbol{b_l}\cdot\partial_t\boldsymbol{{v_B}_l}\,,
\label{eq:theta-subrel}
\end{align}
with $\boldsymbol{b_l}$ the unit field line direction vector after coarse-graining on scale $l$. Coarse-graining does not formally distribute itself over products of fields, but we ignore the coupling between scales that would result from a more formal procedure, and simply coarse-grain all quantities. On scale $l$, the above gradients are of order of magnitude ${\delta u}_l/l$ where $\delta u_l$ henceforth denotes the characteristic root mean squared increment of $u_E$ on scale $l$. The acceleration term differs from the other two by a power of $\delta v_{\rm c}/v$. Consequently, it provides a sub-dominant contribution in sub-relativistic turbulence, unless $v<\delta v_{\rm c}$.  

The dynamical system that couples the evolution of momentum and pitch-angle cosine then takes the  form 
\begin{align}
\frac{{\rm d}}{{\rm d}t} p&\,=\,-\frac{c}{v}p\mu \,{a_\parallel}_l - p{\mu}^2\,{\Theta_\parallel}_l - \frac{1}{2}p\left(1-{\mu}^2\right)\,{\Theta_\perp}_l\,,\nonumber\\
\frac{{\rm d}}{{\rm d}t}\mu&\,=\, -\frac{c}{v}(1-\mu^2)\,{a_\parallel}_l -\mu(1-\mu^2)\,{\Theta_\parallel}_l + \frac{1}{2}\mu(1-\mu^2)\,{\Theta_\perp}_l\, \nonumber\\
&\quad\quad + \xi_\mu.
\label{eq:pp-interm}
\end{align}

Leaving aside for now the $c/v$ dependence in front of the acceleration term in the limit $v\ll c$, to which we will return immediately after Eq.~(\ref{eq:Dp-interm}), the equation of momentum evolution is of the form $\dot p \sim \Gamma_l\, p$, pointing to systematic energization (with exponential increase of $p$) for as long as the random quantity $\Gamma_l$ remains approximately constant. From this system, we seek to derive the transport coefficients 
\begin{align}
A_p\equiv\frac{\langle\Delta p\rangle}{\Delta t}& \,=\, \underset{\Delta t\rightarrow +\infty}{\rm lim}\frac{1}{\Delta t}\int_0^{\Delta t}{\rm d}t_1 \,\left\langle \frac{{\rm d}p}{{\rm d}t_1}\right\rangle\,\nonumber\\
D_{pp}\equiv\frac{\langle\Delta p^2\rangle}{\Delta t}& \,=\, \underset{\Delta t\rightarrow +\infty}{\rm lim}\frac{1}{\Delta t}\int_0^{\Delta t}{\rm d}t_1{\rm d}t_2 \,\left\langle \frac{{\rm d}p}{{\rm d}t_1}\frac{{\rm d}p}{{\rm d}t_2}\right\rangle\,.
\label{eq:interm-coeff}
\end{align}
Averages are taken here over the statistics of all random fields.
This calculation implicitly assumes that the average energy gain per interaction is smaller than unity. It therefore makes sense to calculate these two moments to lowest order in the amplitude of the random forces. 

For a stochastic process of the above form, with multiplicative noise $\propto p$, it is more convenient to use $q\equiv \ln p$ as a variable, with corresponding advection $A_q \equiv\langle\Delta q\rangle/\Delta t$ and diffusion $D_{qq}\equiv\langle\Delta q^2\rangle/\Delta t$ coefficients.  The distribution $n_p\equiv 4\pi p^2 f(p,\,t)$, i.e. the number density of particles per momentum interval, written here in terms of the distribution function $f(p,\,t)$, then obeys the Fokker-Planck equation 
\begin{equation}
    \frac{\partial}{\partial t}n_p \,=\, -\frac{\partial}{\partial p}(A_p n_p) + \frac{\partial^2}{\partial p^2}(D_{pp} n_p)
    \label{eq:FPnp}
\end{equation}
with the correspondence $A_p = p(A_q + D_{qq})$ and $D_{pp} = p^2D_{qq}$. The diffusion coefficient $D_{qq}$ enters the expression for $A_p$ because of the momentum dependence of the energization rate in Eq.~(\ref{eq:interm-coeff}), which induces a net drift in momentum (noise-induced drift of stochastic processes). It is more convenient to calculate $A_q$ and $D_{qq}$, for which the integrand does not depend on $p$, then use the above relations to obtain $A_p$. We do so in the following but write down only the contribution $A'_p \equiv p A_q$ for simplicity. 
As we show further that $A'_p \simeq D_{pp}/p$ up to a coefficient of order unity,   the difference between $A'_p$ and $A_p$ will not be of consequence.

Close observation of Eq.~(\ref{eq:pp-interm}) reveals that the coefficient of net advection $A'_p$ does not vanish trivially, even if the mean value of each gradient vanishes, e.g. $\langle {\Theta_\parallel}_l\rangle=0$, because the evolution of $\mu$ is influenced by the random forces, and hence correlated with these forces, even if $\xi_\mu$ represents by itself an independent process. Further inspection reveals, in fact, that the advection term is systematically positive,  indicating net energy gain. Consider for instance the first term $\propto {\Theta_\parallel}_l$. Assuming ${\Theta_\parallel}_l>0$, the particle loses energy while in the structure; simultaneously, however, $\vert\mu\vert$ decreases under the influence of ${\Theta_\parallel}_l$, implying that the amount of energy lost is less than what would have been calculated in the limit of constant $\mu$. Conversely, if ${\Theta_\parallel}_l<0$, the particle gains more energy than anticipated for constant $\mu$, because ${\Theta_\parallel}_l$ now pushes $\vert\mu\vert$ to larger values. Summing these two contributions for ${\Theta_\parallel}_l<0$ and ${\Theta_\parallel}_l>0$ results in a net energy gain. The same exercise can be repeated for the other terms, with the same conclusion. 

This result can be understood in a more general way in the laboratory frame. Consider a particle undergoing a random walk in momentum space through interactions with randomly oriented electric fields $\boldsymbol{\delta E}$. If the orientation of these fields is truly random, then $\langle \dot p\rangle=0$ to first order in $\boldsymbol{\delta E}$. However, because $\boldsymbol{\delta E}$ also impacts the momentum in such a way as to focus the spatial component of the momentum toward the direction that maximizes the energy gain (i.e. toward $q \boldsymbol{\delta E}$), particles spend more time oriented in that direction than in the opposite. Notwithstanding the contribution of the magnetic part of the Lorentz force, this leads to a net energy gain of order $\boldsymbol{\delta E}^2$. This effect is captured here as $A_p \propto \langle {\Theta_\parallel}_l^2\rangle$. Formally, this effect also represents a noise-induced drift of the overall stochastic process. 

In detail, we use Eq.~(\ref{eq:stat-gamma}) to describe the statistical properties of the random fields, Eq.~(\ref{eq:stat-mu}) to describe that of $\mu_{\{\xi\}}$ (the perpendicular components having been integrated out), with $\nu_{\rm s}<\nu_{\rm f}$ per the assumption $l<\lambda_\parallel$, and we integrate the equation for $\dot p$, taking into account the evolution in time of $\mu$ for $a_\parallel$, or $\mu^2$ for the other terms $\Theta_\parallel$ and $\Theta_\perp$. To order $\langle \Gamma_l^2\rangle$, this gives
\begin{align}
    A'_{p\,\vert l} &\,\underset{\Delta t\rightarrow +\infty}{\simeq}\, \frac{p}{\Delta t}\,\int_0^{\Delta t}{\rm d}t_2\,\int_0^{t_2}{\rm d}t_1 \,\biggl\{\nonumber\\
    &\quad \,\,\,\frac{c^2}{v^2}\left\langle \left(1-\mu_{\{\xi\}}(t_1)^2\right){a_\parallel}_l(t_1){a_\parallel}_l(t_2) \right\rangle\,\nonumber\\
    & \quad + 2\left\langle\mu_{\{\xi\}}(t_1)^2 \left(1-\mu_{\{\xi\}}(t_1)^2\right){\Theta_\parallel}_l(t_1){\Theta_\parallel}_l(t_2) \right\rangle
    \nonumber\\
     & \quad + \frac{1}{2}\left\langle\mu_{\{\xi\}}(t_1)^2 \left(1-\mu_{\{\xi\}}(t_1)^2\right){\Theta_\perp}_l(t_1){\Theta_\perp}_l(t_2) \right\rangle\biggr\}
    \nonumber\\
    &\,\simeq\, p\, \frac{\nu_{\rm f}}{\nu_{\rm f}^2+\omega_{\rm f}^2}\,\left\{ \frac{2}{3} \frac{c^2}{v^2}\langle {a_\parallel}_l^2\rangle + \frac{4}{15}\langle {\Theta_\parallel}_l^2\rangle + \frac{1}{15}\langle {\Theta_\perp}_l^2\rangle\right\}\,.
    \label{eq:Ap-interm}
\end{align}
The terms in curly brackets quantify the rate of advection for each term, while the first prefactor on the last line provides the characteristic response of a damped oscillator (the limit $\nu_{\rm s}\ll\nu_{\rm f}$ has been taken here). The contribution $\propto \langle {a_\parallel}_l^2\rangle$ assumes that $v$ is not much smaller than $c$. Its contribution as $v/c\rightarrow 0$ will be revised further below.

In the original Fermi picture, in which particles undergo discrete, point-like interactions with moving scattering centers, net advection is commonly interpreted as the result of more frequent head-on than tail-on collisions. In reality, the dominant contribution comes from a genuinely relativistic effect similar to the above. A Lorentz transform back and forth to the scattering center frame indeed connects the incoming and outgoing energies, respectively $\epsilon_{\rm in}$ and $\epsilon_{\rm out}$ in the laboratory frame, through $\epsilon_{\rm out}/\epsilon_{\rm in} = \gamma_E^2 \left(1 + \boldsymbol{v_E}\cdot\boldsymbol{v_{\rm out}'}/c^2\right)\left(1-\boldsymbol{v_E}\cdot\boldsymbol{v_{\rm in}}/c^2\right)$. If the dot products are zero on average, one obtains $\epsilon_{\rm out}/\epsilon_{\rm in} - 1 = \gamma_E^2 - 1 \simeq v_E^2/c^2$, a net energy gain indeed. The head-on/tail-on argument results from weighing the Lorentz transform by the flux of incoming particles, assuming an isotropic pre-interaction distribution function, which provides a contribution a factor 3 smaller than that of the above relativistic boost. Now, this Lorentz transform provides the analog of our calculation in \Rco. If one were to perform the calculation in the laboratory frame, in which the electric fields do not vanish but are everywhere related to magnetic fluctuations $\boldsymbol{\delta B}$ through $\boldsymbol{\delta E}=-\boldsymbol{v_E}\times\boldsymbol{\delta B}/c$, one would compute the global energy gain as the sum of a number of energy increments of order $O(\delta E^2)$ per interaction, for the same reasons as discussed previously, with a global value $O(v_E^2/c^2)$~\footnote{From a purely kinematic point of view, recalling that energy gain is positive when momentum transfer is aligned with $\boldsymbol{v_E}$, one can also view the relativistic contribution to advection (i.e., the term $\gamma_E^2$) as the consequence of relativistic focusing along $\boldsymbol{v_E}$ of a population of particles that is isotropic in the rest frame of the scattering center.}. 

Going back to the present calculation, the diffusion coefficient can be evaluated to lowest order in the squared amplitudes of the random forces by direct integration of the equation of evolution, with substitution $\mu(t) \rightarrow \mu_{\{\xi\}}(t)$. This gives
\begin{align}
    D_{pp\,\vert l}&\,=\,p^2\frac{\nu_{\rm f}}{\nu_{\rm f}^2+\omega_{\rm f}^2}\left\{ \frac{2}{3} \frac{c^2}{v^2}\langle {a_\parallel}_l^2\rangle + \frac{2}{5}\langle {\Theta_\parallel}_l^2\rangle + \frac{4}{15}\langle {\Theta_\perp}_l^2\rangle\right\}\,.
    \label{eq:Dp-interm}
\end{align}
Comparison with Eq.~(\ref{eq:Ap-interm}) shows that up to prefactors of the order of unity, $A'_{p\,\vert l}\simeq D_{pp\,\vert l}/p$ (and therefore $A_{p\,\vert l}\simeq D_{pp\,\vert l}/p$ as well), meaning that this regime obeys an approximate -- but not exact -- regime of detailed balance. For reference, detailed balance corresponds to the assumption $A_p = \left(2D_{pp} + p\,{\rm d}D_{pp}/{\rm d}p\right)/p$, for which the Fokker-Planck equation takes a purely diffusive form (in isotropic 3D momentum space), $\partial_t f(p,\,t) = \boldsymbol{\nabla_p}\cdot D_{pp}\,\boldsymbol{\nabla_p} f(p,t) = p^{-2}\partial_p\left[p^2D_{pp}\partial_p f(p,\,t)\right]$. This equation is commonly used in phenomenological applications of stochastic acceleration, for simplicity and due to a  lack of better knowledge. However, the underlying assumption is not supported by fundamental principles~\cite{1997APh.....6..271O}. In the original Fermi picture, for instance, detailed balance 
holds only if the cross-section is velocity-independent~\cite{2005PPCF...47B.667D}. In kinetic numerical simulations of trans-relativistic turbulence (meaning $\delta v_{\rm c}$ not much smaller than $c$), the measured advection and diffusion coefficients do not obey detailed balance either~\cite{2020ApJ...893L...7W,2025arXiv250203042W}. This last result provides a special motivation to investigate the origin and nature of advection.

The contribution of the acceleration term to the advection and diffusion coefficients must be corrected in the sub-relativistic limit $v\ll c$, as announced earlier. The corresponding contribution is then best rewritten for the variables $p_\parallel \equiv p \mu$ and $p_\perp \equiv p\sqrt{1-\mu^2}$, giving $\dot p_\parallel \simeq - mc a_\parallel$, $\dot p_\perp \simeq 0$, which describes the motion of a sub-relativistic particle subject to a stochastic gravity force $-c\,a_\parallel$, recalling that $a_\parallel$ has been rescaled by $c$ for dimensional consistency. Performing the integrals in Eq.~(\ref{eq:interm-coeff}) then yields $A_p \simeq 0$ and $D_{pp} \simeq 2 m^2 c^2\nu_{\rm f}(\nu_{\rm f}^2 + \omega_{\rm f}^2)^{-1} \langle {a_\parallel}_l^2\rangle $. In short, the contribution of acceleration to advection effectively vanishes in the sub-relativistic limit, while the expression of $D_{pp}$ given in Eq.~(\ref{eq:Dp-interm}) remains valid.

The numerical prefactors that enter Eqs.~(\ref{eq:Ap-interm}) and (\ref{eq:Dp-interm}) depend on the assumptions made regarding the statistical properties of $\mu_{\{\xi\}}$. In particular, we have complemented Eq.~(\ref{eq:stat-mu}) with 
\begin{align}
    \langle \mu_{\{\xi\}}(t_1)^2\mu_{\{\xi\}}(t_2)^2 \rangle &\,\equiv\, \frac{1}{5}\,e^{-\nu_{\rm s}\vert t_2-t_1\vert}\,+\,  \frac{1}{9}\,\left(1-e^{-\nu_{\rm s}\vert t_2-t_1\vert}\right)\,,
    \label{eq:stat-mu2}
\end{align}
where $1/5$ corresponds to the average of $\mu^4$ uniformly distributed over the interval $[-1,+1]$. Given that detailed balance is not exactly verified, and that these prefactors are somewhat model dependent, we ignore them in the following and note that all prefactors will be of the order of $0.1\ldots 0.5$ as in the present case. 

The response $\propto \nu_{\rm f}/(\nu_{\rm f}^2+\omega_{\rm f}^2)$ implies that the transport coefficients vanish in both limits $\nu_{\rm f}\rightarrow0$ and $\nu_{\rm f}\rightarrow+\infty$. In the former limit, the contributions average out due to the inversion of polarity at rate $\omega_{\rm f}$; in the latter, it is because the interaction time then shrinks to zero. For the present range of scales $l<\lambda_\parallel$, $\nu_{\rm f}$ can be approximated by the maximum of $v\mu/l$ (inverse of crossing time in the parallel direction) and $\delta u_l/l$, the mean eddy turn-around time on scale $l$. Strictly speaking, rather than $v\mu$, one should consider the relative velocity $\Delta v$ between the particle parallel velocity and the parallel velocity of the structure, because structures can move relative to the frame \Rco\ \footnote{In the particular cases of $a_\parallel$ and $\Theta_\parallel$, this issue becomes subtle, as obviously a particle surfing on a bend of the field line with zero relative velocity would not probe any velocity shear and therefore would not experience any acceleration. On a formal level, both parallel shear and curvature drift originate from the same term $p_1'{u_E}^\beta\,{\rm d}b_\beta/{\rm d}\tau$ where ${\rm d}/{\rm d}\tau \,\equiv\, (p^\gamma/m)\partial_\gamma$ denotes a derivative taken along the particle trajectory,  see also Eq.~(A18) of Ref.~\cite{2021PhRvD.104f3020L}. If ${\rm d}b_\beta/{\rm d}\tau=0$, the contribution vanishes as it should. This case remains an exception, which does not apply to other force terms.}. At resonance $\Delta v=0$ (analogous to the Landau resonance of transit-time damping), or in the low-velocity limit $v < \delta u_l$, particles are effectively trapped in the structure, so that the decorrelation rate becomes limited by the structure turn-around time. We will return to this point in Sec.~\ref{sec:disc} and consider here the generic case on small scales, for which $\Delta v \gtrsim \delta u_l$, i.e. $\nu_{\rm f}\simeq v/l$. Regarding the frequency $\omega_{\rm f}$, one likely has $\omega_{\rm f}\simeq\nu_{\rm f}\sim v/l$ for similar reasons, and hence (up to the scaling $c^2/v^2$ for the advection term, see above)
\begin{align}
    \frac{A_{p\,\vert l}}{p}\,\sim\,\frac{D_{pp\,\vert l}}{p^2}\,\sim\, \left\{ \frac{c^2}{v^2}\langle {a_\parallel}_l^2\rangle + \langle {\Theta_\parallel}_l^2\rangle + \langle {\Theta_\perp}_l^2\rangle\right\}\,\frac{l}{v}\,.
    \label{eq:D-interm-2}
\end{align}
We have neglected here the possibility that the amplitude of the mode is large enough to trap the particle through its influence over $\mu$, in agreement with our approach to the transport coefficients. This may occur, in particular, for low-$\mu$ particles in magnetic bottle configurations provided the magnetic perturbation $\delta B_l/B \gtrsim \mu^2/(1-\mu^2)$, as originally proposed by Fermi himself~\cite{1954ApJ...119....1F}, and recently investigated in~\cite{2023ApJ...956...63L}. The above result can however be generalized to that case in a straightforward manner. For efficient trapping, $\nu_{\rm f}\sim \delta u_l/l$, in which case the above formula must be modified by $v\rightarrow \delta u_l$ in the last prefactor.

To determine the scaling of the above contributions with $l$, assume standard MHD turbulent scalings $\delta u_l\propto \delta v_{\rm c}\,(l/\ell_{\rm c})^{(q-1)/2}$, with $q$ the index of the power spectrum of turbulent velocity fluctuations and $\delta v_{\rm c}$ the characteristic velocity of $v_E$ on the outer scale, e.g. $\delta v_{\rm c} \simeq v_{\rm A}\delta B/B_0$ for Alfvénic turbulence. Both ${\Theta_\parallel}_l$ and ${\Theta_\perp}_l$ scale as a velocity gradient $\delta u_l/l$. In order of magnitude, the contribution on scale $l$ scales as 
\begin{align}
    {\Theta_\parallel}_l,\,{\Theta_\perp}_l:\quad\quad \frac{A_{p\,\vert l}}{p}\,\sim\,\frac{D_{pp\,\vert l}}{p^2}\,\sim\, \frac{\delta v_{\rm c}^2}{v\,\ell_{\rm c}} \left(\frac{l}{\ell_{\rm c}}\right)^{q-2}\,.
    \label{eq:sc-theta-interm}
\end{align}
If $q\simeq 5/3$, these contributions scale as $l^{-1/3}$, indicating that small scales $l\sim r_{\rm g}$ dominate.

The acceleration $a_\parallel$ generically scales as $(\delta u_l/c)\delta u_l/l$, see Eq.(\ref{eq:theta-subrel}), so that its contribution to the diffusion coefficient scales as $D_{pp\,\vert l}/p^2 \propto (\delta v_{\rm c}/v)^2 \delta v_{\rm c}^2/(c \ell_{\rm c}) (l/\ell_{\rm c})^{2q-3}$. It is therefore a factor $(\delta v_{\rm c}/v)^2 (l/\ell_{\rm c})^{q-1}$ smaller than the contributions of $\Theta_\parallel$ and $\Theta_\perp$. It  gains importance, however, in the sub-relativistic limit and could eventually become dominant in the low-velocity regime $v \lesssim \delta u_l$, where particles become trapped in the structure for an eddy turn-around time.

The diffusion and advection coefficients for $\Theta_\parallel$, which corresponds here to dynamic field line curvature energization (or curvature drift acceleration), have not been presented before. This curvature contribution was discussed in Ref.~\cite{2021PhRvD.104f3020L} and modeled in more detail in Ref.~\cite{2022PhRvL.129u5101L} using its full probability distribution function, assuming a momentum jump of amplitude ${\Theta_\parallel}_l$ over timescale $l/v$, with a dominant contribution on scale $l\sim r_{\rm g}$, in agreement therefore with the above trend. That same reference showed indeed that in large amplitude turbulence $\delta B/B_0 \gtrsim 1$, the power-law tails of the pdf implied that the Fokker-Planck treatment was not adequate, as the stochastic process takes the form of a truncated Lévy flight. If, at $\delta B/B_0\lesssim 1$, the tails of the pdf become softer, which is indeed suggested by direct statistical measurements to be reported in a companion paper, then Eq.~(\ref{eq:D-interm-2}) provides the corresponding diffusion coefficient. In the context of turbulence, evidence for curvature drift acceleration has been reported in several numerical experiments, see for instance \cite{2018PhRvL.121x5101A,2020ApJ...894..136T,2022PhRvD.106b3028B}.

The result $D_{pp}\sim p^2 (r_{\rm g}/\ell_{\rm c})^{-1/3} \delta v_{\rm c}^2/(\ell_{\rm c} v)$ for ${\Theta_\perp}_l$ with $q=5/3$ matches the quasilinear estimate for transit-time damping. This is expected in so far as transit-time damping can also be described as particle acceleration through interaction with magnetic mirrors of finite amplitude~\cite{1981A&A....97..259A}.

\subsection{Contribution from large scales, for $\delta B/B_0 \gtrsim 1$}\label{sec:large}
We now consider the limit $l\gtrsim \lambda_\parallel$ and further assume that $\delta B/B_0\gtrsim 1$ so that transport can be described as approximately isotropic, i.e. with roughly equal parallel and perpendicular diffusion coefficients. Correspondingly, we write $\lambda_\parallel \rightarrow \lambda_{\rm s}$ for the scattering mean free path. For reference, $\lambda_{\rm s} \equiv v/\nu_{\rm s}$; alternatively one may use the spatial diffusion coefficient $\kappa_{\rm s}$ induced by pitch-angle scattering, $\kappa_{\rm s}\equiv \lambda_{\rm s}v\equiv v^2/\nu_{\rm s}$ (omitting $1/3$ prefactors for simplicity). We neglect any possible eddy anisotropy, but we return to this issue in Sec.~\ref{sec:aniso}. 

In the present limit, the direction of the magnetic field does not play any role, and we may assume that the particle scatters on magnetic perturbations carried by the background plasma, so that $\boldsymbol{v_E}\simeq \boldsymbol{v_{\rm p}}$. The random dynamical system takes the form
\begin{align}
    \dot p &\,=\, -p\times\biggl\{\frac{c}{v}\mu_1\, {a_1}_l + \mu_1^2\,{\Theta_1}_l + \mu_2^2\, {\Theta_2}_l + \mu_2\mu_3\,{\sigma_{23}}_l  \nonumber\\
    &\quad\quad  + \mu_1\mu_2\, ({\sigma_{12}}_l+{\sigma_{21}}_l) + (2\leftrightarrow 3)\biggr\}\nonumber\\
    \dot \mu_1 &\,=\, -\frac{c}{v}(1-\mu_1^2)\,{a_1}_l - \mu_1(1-\mu_1^2)\,{\Theta_1}_l + \mu_1\mu_2^2 \,{\Theta_2}_l  \nonumber\\
    &\quad\quad - \mu_2(1-\mu_1^2)\,{\sigma_{12}}_l + \mu_1^2\mu_2{\sigma_{21}}_l\nonumber\\
    &\quad\quad +\mu_1\mu_2\mu_3\,{\sigma_{23}}_l + \xi_{\mu_1} + (2\leftrightarrow 3)\nonumber\\
    \dot \mu_2 &\,=\, \mu_1^2\mu_2\, {\Theta_1}_l -\mu_2(1-\mu_2^2)\,{\Theta_2}_l \nonumber\\
    &\quad\quad + \mu_1\mu_2^2\,{\sigma_{12}}_l -\mu_1(1-\mu_2^2){\sigma_{21}}_l\nonumber\\
    &\quad\quad -\mu_3(1-\mu_2^2)\,{\sigma_{23}}_l +\xi_{\mu_2}+ (1\leftrightarrow 3)\,.
    \label{eq:sys-largel}
\end{align}
The term for $\dot \mu_3$ can be similarly derived from Eq.~(\ref{eq:dyn_mu1_0}). The above system shows that the angular variables along each direction decouple one from the other in their respective contributions to acceleration and compression, meaning that one can solve for $p$ and $\mu_1$ only to obtain the contributions of ${a_1}_l$ or ${\Theta_1}_l$, irrespective of the evolution of $\mu_2$ and $\mu_3$. The shear contribution ${\sigma_{ij}}_l$, however, requires us to track $p$, $\mu_i$ and $\mu_j$. 

Contrary to the previous regime $l<\lambda_\parallel$, angular diffusion now plays an important role. The time needed for a particle to explore (and exit from) a structure becomes $\sim l^2/(\lambda_s v)$. Correspondingly, the decorrelation frequency will be given by
\begin{align}
    \nu_{\rm f}&\,\simeq\,{\rm max}\left\{\frac{\lambda_{\rm s}v}{l^2},\,\,\frac{\delta u_l}{l}\right\}\,.
    \label{eq:nuf-large}
\end{align}
Regarding the frequency $\omega_{\rm f}$, it can be written as
\begin{align}
    \omega_{\rm f}&\,\simeq\,{\rm max}\left\{k_{\rm f0}^2\lambda_{\rm s}v,\,\,\omega_{\rm f0}\right\}\,,
    \label{eq:omegaf-large}
\end{align}
with $k_{\rm f0}\sim l^{-1}$. The first term describes the diffusive transport in a spatial structure whose polarity alternates on scale $\sim l$, while the second corresponds to the proper oscillating frequency, $\omega_{\rm f0}\sim v_{\rm f0}/l$, where $v_{\rm f0}$ is some characteristic phase velocity. Overall, this suggests that the hierarchy between $\nu_{\rm f}$ and $\omega_{\rm f}$ is rather restricted, either because $v_0 \sim \delta u_l$, or because $v_0$ is given by a characteristic sound or Alfvén speed and $\delta u_l$ scales slowly with $l$. In the following, we will use the short-hand notation 
\begin{equation}
    R_{\omega/\nu}\equiv \frac{1}{1+\omega_{\rm f}^2/\nu_{\rm f}^2}\,,
    \label{eq:def-Ron}
\end{equation} 
and retain it as a general prefactor in the expressions that follow.

The global influence of scattering on the calculation of the advection and diffusion coefficients can be read from the system Eq.~(\ref{eq:sys-largel}), recalling the procedure adopted in Sec.~\ref{sec:interm} to calculate these coefficients. The contribution of ${a_1}_l$, for instance, brings a term $\langle {\mu_{\{\xi\}}}_1^2(t_1)\rangle$ in the time integral for $A_p$, which therefore takes a non-zero value even if $\nu_{\rm s}\rightarrow +\infty$, while the calculation of the diffusion coefficient involves a term $\langle {\mu_{\{\xi\}}}_1(t_1){\mu_{\{\xi\}}}_1(t_2)\rangle$ that vanishes in that limit. The shear contribution contains a similar unequal time correlation function in both advection and diffusion terms, indicating that both coefficients vanish in the limit $\nu_{\rm s}\rightarrow +\infty$.  On the contrary, efficient pitch-angle scattering does not affect advection or diffusion in compressive modes (${\Theta_i}_l$) in a significant way, because the expression for these coefficients itself depends weakly on the time correlation function of ${\mu_{\{\xi\}}}_i$. 

Particle acceleration through scattering in dynamical large-scale compressive fluctuations is also termed  ``magnetic pumping'' in some contexts~\cite{2017ApJ...850L..28L,2020NatCo..11.2942L,2023ApJ...947...89L}. While previous references have discussed this mechanism through a linearized transport equation, assuming conservation of adiabatic invariants, the above treatment is based on an {\it ab initio} microscopic formulation. The general scaling of the diffusion coefficient derived below in Eq.~(\ref{eq:AD-aT-large}),  $D_{pp}\propto \nu_{\rm f}/(\nu_{\rm f}^2+\omega_{\rm f}^2)$ matches that obtained in Ref.~\cite{2017ApJ...850L..28L}. Differences in the final scalings of $D_{pp}$ between the present and earlier studies then relate to the choice of $\nu_{\rm f}$ and $\omega_{\rm f}$, therefore on the model of transport through the structures and the conditions at hand.

In the present context, it becomes important to account for the possibility of turbulent diffusion, which ensures finite spatial transport even as $\nu_{\rm s}\rightarrow+\infty$. Recalling our earlier discussion, the velocity field of structures of scales $\lesssim l$ is non zero, but random, in the frame \Rco. On scale $l$, the characteristic diffusion coefficient associated with turbulent motion is of the order of $l\delta u_l$ up to a numerical prefactor that we neglect. Turbulent diffusion can be included in the evolution of the pitch angle variables with the following modification
\begin{align}
    \langle {\mu_{\{\xi\}}}_i \rangle &\,=\,0\nonumber\\
    \langle {\mu_{\{\xi\}}}_i(t_1){\mu_{\{\xi\}}}_j(t_2) \rangle &\,=\,\frac{1}{3}\delta_{ij}\biggl[(1-x_{\rm t}^2)\,e^{-\nu_{\rm s}\vert t_1-t_2\vert}\nonumber\\
    &\quad \quad\quad \,+\, x_{\rm t}^2\,e^{-\nu_{\rm t}\vert t_1-t_2\vert}\biggr]\,
    \label{eq:stat-mu-turb}
\end{align}
where $\nu_{\rm t} \sim \delta u_l/l$ provides the rate of decorrelation of the direction of motion of an eddy on scale $l$ or less, and $x_{\rm t}^2 \equiv  \delta u_l^2/(v^2 + \delta u_l^2)$ is expressed in terms of the mean eddy velocity $\delta u_l$ on scale $l$. The above statistics can be directly recovered by relating the particle velocity in \Rco\ to that in the frame in which the turbulent velocity vanishes exactly (assuming a Galilean transform), then converting to pitch angle variables and taking the statistical moments.

Turbulent diffusion contributes on all scales, including the intermediate scales ($l<\lambda_{\rm s}$) discussed in Sec.~\ref{sec:interm}. With respect to the results obtained in that section, the inclusion of turbulent diffusion brings in an additional  contribution from the shear, which was eliminated by the gyro-averaging procedure. However, the gyro-average should formally be taken in the frame of small-scale eddies (on scale $\sim r_{\rm g}$) which retain a non-zero random velocity relative to the frame \Rco\ on scale $l$. Turbulent diffusion then restores a net transverse motion with a correlation time that is significantly longer than the gyro-time. Nevertheless, a direct calculation of the corresponding contribution shows that it grows with scale $l$, and hence that it is  dominated by the (present) range of scales $l\gtrsim\lambda_{\rm s}$.

The rates verify the hierarchy $\nu_{\rm s} > \nu_{\rm f} \geq \nu_{\rm t}$, since
\begin{equation}
    \frac{\lambda_{\rm s}v}{l^2}\,\frac{1}{\nu_{\rm s}}\,=\, \frac{\lambda_{\rm s}^2}{l^2}\, <\,1\,,\quad
    \frac{\delta u_l}{l}\,\frac{1}{\nu_{\rm s}} \,=\, \frac{\nu_{\rm t}}{\nu_{\rm s}} \,=\, \frac{\delta u_l}{v}\,\frac{\lambda_{\rm s}}{l}\,<\,1\,.
\end{equation}
The first two imply that $\nu_{\rm f}$ is always smaller than $\nu_{\rm s}$, while the third implies that angular decorrelation through turbulent diffusion is always slower than that due to pitch-angle scattering in the range of scales $l\gtrsim\lambda_{\rm s}$. With this preamble, we can now calculate the advection and diffusion coefficients as in the previous section, first for the acceleration and compressive terms, omitting as before the numerical prefactors. To simplify slightly the expressions, we further retain only the leading terms in $x_{\rm t}^2$ and $\nu_{\rm f}/\nu_{\rm s}$. This gives
\begin{alignat}{2}
    {a_1}_l:&\, \frac{A'_{p\,\vert l}}{p} &&\,\sim\, \frac{c^2}{v^2}\langle {a_1}_l^2\rangle\,\frac{\nu_{\rm f}}{\nu_{\rm f}^2 + \omega_{\rm f}^2}\nonumber\\
    {a_1}_l:&\, \frac{D_{pp\,\vert l}}{p^2}  &&\,\sim\, \frac{c^2}{v^2}\langle {a_1}_l^2\rangle\,\biggl[\frac{\nu_{\rm f}+\nu_{\rm s}}{(\nu_{\rm f}+\nu_{\rm s})^2 + \omega_{\rm f}^2} 
    \nonumber\\\  & &&\quad\quad 
    + x_{\rm t}^2 \frac{\nu_{\rm f}+\nu_{\rm t}}{(\nu_{\rm f}+\nu_{\rm t})^2 + \omega_{\rm f}^2}\biggr]\nonumber\\
    {\Theta_1}_l:&\, \frac{A'_{p\,\vert l}}{p} &&\,\sim\, \langle {\Theta_1}_l^2\rangle\,\frac{\nu_{\rm f}}{\nu_{\rm f}^2 + \omega_{\rm f}^2}\nonumber\\
    {\Theta_1}_l:&\, \frac{D_{pp\,\vert l}}{p^2} &&\,\sim\, \langle {\Theta_1}_l^2\rangle\,\frac{\nu_{\rm f}}{\nu_{\rm f}^2 + \omega_{\rm f}^2}\,.
    \label{eq:AD-aT-large}
\end{alignat}

To solve for the shear contribution $\sigma_{12}$, we track the evolution of $p$, $p_1 = p\mu_1$ and $p_2= p\mu_2$, because it provides a simpler system that decouples the two variables: 
\begin{align}
    \dot p &\,=\, -\frac{1}{p}p_1p_2 \sigma_{12} + \ldots \nonumber\\
    \dot p_1 &\,=\, -p_2 \sigma_{12} + \ldots\nonumber\\
    \dot p_2 &\,=\, 0\,,
    \label{eq:sh-syst-large}
\end{align}
Noting that $p$ is kept constant in the integration over time in the calculations of $A_p$ and $D_{pp}$, this provides a solution to the same order of approximation. Furthermore, $\mu_1\mu_2$ can be regarded as an angular variable decorrelating at rate $\nu_{\rm s}$, yet with mean squared value $\langle\mu_1^2\mu_2^2\rangle =1/15$ instead of $1/3$. Modifying Eq.~(\ref{eq:stat-mu-turb}) in consequence, we obtain
\begin{alignat}{2}
    {\sigma_{12}}_l: &\, \frac{A'_{p\,\vert l}}{p} &&\,\sim\, \langle {\sigma_{12}}_l^2\rangle\,\biggl[\frac{\nu_{\rm f}+\nu_{\rm s}}{(\nu_{\rm f}+\nu_{\rm s})^2 + \omega_{\rm f}^2} 
    \nonumber\\ & &&\quad\quad 
    + x_{\rm t}^2 \frac{\nu_{\rm f}+\nu_{\rm t}}{(\nu_{\rm f}+\nu_{\rm t})^2 + \omega_{\rm f}^2}\biggr]\nonumber\\
    {\sigma_{12}}_l: &\, \frac{D_{pp\,\vert l}}{p^2} &&\,\sim\, \langle {\sigma_{12}}_l^2\rangle\,\biggl[\frac{\nu_{\rm f}+\nu_{\rm s}}{(\nu_{\rm f}+\nu_{\rm s})^2 + \omega_{\rm f}^2} 
    \nonumber\\ & &&\quad\quad 
    + x_{\rm t}^2 \frac{\nu_{\rm f}+\nu_{\rm t}}{(\nu_{\rm f}+\nu_{\rm t})^2 + \omega_{\rm f}^2}\biggr]\nonumber\\
    \label{eq:AD-sig-large}
\end{alignat}
Shear acceleration is often considered in the context of large-scale, ordered velocity flows~\cite{1981SvAL....7..352B,2018ApJ...855...31W,2019Galax...7...78R}. The above offers a direct calculation of the advection and diffusion rate in such settings by taking the limit $\nu_{\rm f}\rightarrow 0$, $\omega_{\rm f}\rightarrow 0$. One recovers the usual scaling $A_p/p\sim D_{pp}/p^2\propto 1/\nu_{\rm s}$, plus a contribution due to turbulent diffusion which is usually ignored in such calculations.

As anticipated above (see also \cite{1983ICRC....9..313B}), turbulent transport ensures finite transport coefficients even in the limit $\nu_{\rm s}\rightarrow+\infty$. While particles are trapped in a small-scale eddy by fast pitch-angle scattering, they remain able to explore the velocity gradients on scale $\sim l$ thanks to the relative random motion of these small-scale eddies. Henceforth, we remove the numbering on the acceleration, and the compressive and shear terms, because statistical isotropy of the forces implies that the total contribution to acceleration is three times that of $a_1$, that of compression is three times that of $\Theta_1$ and that of shear is six times that of $\sigma_{12}$. 

The scaling of the advection and diffusion terms as a function of scale can then be derived after specifying the scale $l_\kappa$ at which $\nu_{\rm f}$ transits from one scaling to the other in Eq.~(\ref{eq:nuf-large}). This scale $l_\kappa$ is such that 
\begin{equation}
l_\kappa \delta u_{l_\kappa} = \lambda_{\rm s}v\,, 
\label{eq:lkappa} 
\end{equation}
i.e., on scale $l_\kappa$, the spatial turbulent diffusion coefficient becomes comparable to the spatial diffusion coefficient $\kappa_{\rm s}$ that derives from  pitch-angle scattering. For relativistic particles and sub-relativistic turbulence, this scale is likely larger than $\lambda_{\rm s}$, because $l_\kappa/\lambda_{\rm s} = v/\delta u_{l_\kappa}$; we assume here that this is the case. 

For $\lambda_{\rm s}< l< l_\kappa$, $\nu_{\rm f}$ is given by the first term in the brackets of Eq.~(\ref{eq:nuf-large}), and $\nu_{\rm s}>\nu_{\rm f}\sim v^2/\nu_{\rm s}l^2 > \nu_{\rm t}$, meaning that decorrelation takes place through diffusive escape,  transport being governed by scattering on magnetic perturbations. Consequently, Eqs.~(\ref{eq:AD-aT-large}) and (\ref{eq:AD-sig-large}) can be further approximated as
\begin{alignat}{2}
    \lambda_{\rm s}< l< l_\kappa: & && \nonumber\\
    {a}_l:&\quad \frac{A'_{p\,\vert l}}{p} &&\,\sim\, \frac{c^2}{v^2}\langle a^2_l\rangle\,R_{\omega/\nu}\, \frac{l^2}{v^2}\nu_{\rm s}\nonumber\\
    {a}_l:&\quad \frac{D_{pp\,\vert l}}{p^2}&&\,\sim\, \frac{c^2}{v^2}\langle a^2_l\rangle\,R_{\omega/\nu}\,\left[\frac{1}{\nu_{\rm s}}+ x_t^2\frac{l^2}{v^2}\nu_{\rm s}\right] \nonumber\\
    {\Theta}_l:&\quad  \frac{A'_{p\,\vert l}}{p} &&\,\sim\, \langle \Theta^2_l\rangle\,R_{\omega/\nu}\, \frac{l^2}{v^2}\nu_{\rm s}\nonumber\\
    {\Theta}_l:&\quad \frac{D_{pp\,\vert l}}{p^2}&&\,\sim\, \langle \Theta^2_l\rangle\,R_{\omega/\nu}\, \frac{l^2}{v^2}\nu_{\rm s}\nonumber\\
    {\sigma}_l:&\quad \frac{A'_{p\,\vert l}}{p} &&\,\sim\, \langle \sigma^2_l\rangle\,R_{\omega/\nu}\, \left[\frac{1}{\nu_{\rm s}}+ x_t^2\frac{l^2}{v^2}\nu_{\rm s}\right] \nonumber\\    
    {\sigma}_l:&\quad  \frac{D_{pp\,\vert l}}{p^2}&&\,\sim\, \langle \sigma^2_l\rangle\,R_{\omega/\nu}\, \left[\frac{1}{\nu_{\rm s}}+ x_t^2\frac{l^2}{v^2}\nu_{\rm s}\right]\,.
    \label{eq:AD-large-1}
\end{alignat}
These relations bring forward two types of scaling, which are directly related to the influence of pitch-angle scattering on the acceleration process. In the compressive terms, scattering does not affect acceleration, which therefore proceeds as long as particles are confined in the structure~\footnote{Scattering on magnetic perturbations couples the particles to the plasma, and hence plasma compression implies energization, by virtue of phase space volume conservation.}, i.e. on a timescale $l^2\nu_{\rm s}/v^2\simeq l^2/\kappa_{\rm s}$. By contrast, the accelerating force in a shear flow changes polarity as the particle changes direction, and hence the rate at which this occurs controls the scaling of the transport coefficients. The corresponding brackets contain two terms, the first associated with pitch-angle scattering, the second with turbulent diffusion. The shear contribution $\propto 1/\nu_{\rm s}$ has been obtained previously in Ref.~\cite{2013ApJ...767L..16O} through a quasilinear formulation of the Fokker-Planck equation, assuming diffusive transport in large-scale modes of a Kolmogorov-type cascade. Since $x_t^2 \sim \delta u_l^2/v^2$, the second term is of order $(l\delta u_l)^2/\kappa_{\rm s}^2$ relative to the first, which is small compared to unity for $\lambda_{\rm s}< l < l_\kappa$, but of order unity at $l\simeq l_\kappa$. The contributions scaling as $1/\nu_{\rm s}$ decrease with increasing $l$ and are thus dominated by the small-scale contributions $l\simeq\lambda_{\rm s}$, or even further below for the acceleration term (Sec.~\ref{sec:interm}). All other terms, associated with turbulent transport or compressive modes, increase with increasing scale $l$ and will therefore be dominated by their contribution at $l\sim l_\kappa$, which we now examine.

At $l>l_\kappa$, $\nu_{\rm s}>\nu_{\rm f}\sim \delta u_l/l = \nu_{\rm t}$, which expresses the fact that decorrelation now takes place on an eddy turn-around time, and that particle transport is governed by turbulent diffusion rather than by scattering. Equations (\ref{eq:AD-aT-large}) and (\ref{eq:AD-sig-large}) then become
\begin{alignat}{2}
    l_\kappa\,<\, l: & && \nonumber\\
    {a}_l:&\quad \frac{A'_{p\,\vert l}}{p} &&\,\sim\, \frac{c^2}{v^2}\langle a^2_l\rangle\,R_{\omega/\nu}\, \frac{l}{\delta u_l}\nonumber\\
    {a}_l:&\quad \frac{D_{pp\,\vert l}}{p^2}&&\,\sim\, \frac{c^2}{v^2}\langle a^2_l\rangle\,R_{\omega/\nu}\,\frac{l \delta u_l}{v^2} \nonumber\\
    {\Theta}_l:&\quad  \frac{A'_{p\,\vert l}}{p} &&\,\sim\, \langle \Theta^2_l\rangle\,R_{\omega/\nu}\, \frac{l}{\delta u_l}\nonumber\\
    {\Theta}_l:&\quad \frac{D_{pp\,\vert l}}{p^2}&&\,\sim\, \langle \Theta^2_l\rangle\,R_{\omega/\nu}\, \frac{l}{\delta u_l}\nonumber\\
    {\sigma}_l:&\quad \frac{A'_{p\,\vert l}}{p} &&\,\sim\, \langle \sigma^2_l\rangle\,R_{\omega/\nu}\, \frac{l \delta u_l}{v^2} \nonumber\\    
     {\sigma}_l:&\quad  \frac{D_{pp\,\vert l}}{p^2}&&\,\sim\, \langle \sigma^2_l\rangle\,R_{\omega/\nu}\, \frac{l \delta u_l}{v^2}\,.
    \label{eq:AD-large-2}
\end{alignat}
With respect to Eq.~(\ref{eq:AD-large-1}), the above expressions have been further simplified by neglecting $1/\nu_{\rm s}$ in front of $x_t^2/\nu_{\rm f}$ at $l>l_\kappa$ and writing $x_t^2/\nu_{\rm f} \simeq l \delta u_l/ v^2$. The contributions scaling as $l/\delta u_l$ are those for which the acceleration is bounded by the residence time. In the present range of scales in which turbulent diffusion dominates spatial transport, the diffusive crossing time of the structure is comparable to the turn-around time of the eddy ($l/\delta u_l$). The contributions scaling as $l\delta u_l/v^2$ are those for which the acceleration rate is limited by the effective scattering rate of the particles dictated by turbulent transport. The contribution of compressive modes $\propto l/\delta u_l$ has first been captured in Ref.~\cite{1983ICRC....9..313B} and likewise interpreted as turbulent transport of particles, and it agrees with subsequent studies~\cite{1988SvAL...14..255P,2003ApJ...599.1426C,2006ApJ...638..811C}.

These contributions scale with $l$ as follows: $D_{pp\,\vert l}(a),\, A_{p\,\vert l}(\sigma),\, D_{pp\,\vert l}(\sigma) \propto l^{(3q-5)/2}$, which is thus constant ($q=5/3$) or slowly evolving with $l$; $A_{p\,\vert l}(a),\, A_{p\,\vert l}(\Theta),\, D_{pp\,\vert l}(\Theta) \propto l^{(q-3)/2}$ decrease with scale. 

To briefly summarize this regime of large-scale modes, with $\delta B/B_0\gtrsim 1$, we observe three different types of contributions to the diffusion coefficient. For the compressive modes, the dominant contribution is that of scales $l\gtrsim l_\kappa$, with $D_{pp} \sim \langle \Theta_{l_\kappa}^2\rangle l_\kappa/\delta u_{l_\kappa}$, see Eq.~(\ref{eq:AD-large-2}) above.  For the shear and acceleration parts, two different scales contribute: first, at $l\sim \lambda_{\rm s}$, with $D_{pp} \sim \langle \sigma_{\lambda_{\rm s}}^2\rangle \nu_{\rm s}^{-1}$ (and similarly for the acceleration), see Eq.~(\ref{eq:AD-large-1}); second, at $l\sim l_\kappa$, with $D_{pp} \sim \langle \sigma_{l_\kappa}^2\rangle l_\kappa \delta u_{l_\kappa}/v^2$, see Eq.~(\ref{eq:AD-large-2}). Note that $l_\kappa$ is generically larger than $\lambda_{\rm s}$ in sub-relativistic turbulence, but for the above scalings to make sense, it must also be smaller than $\ell_{\rm c}$, which provides an upper bound on the particle rigidity (gyroradius), given that $l_\kappa/\ell_{\rm c} = (\kappa_{\rm s}/\ell_{\rm c}\delta v_{\rm c})^{3/4}$, assuming $q=5/3$.

Which contribution dominates depends on the nature of the turbulence (compressive versus noncompressive), on the characteristic eddy velocity on the outer scale and to some degree on the particle gyroradius. These issues will be discussed further and put into a broader context, along with the contributions of the intermediate scales $l\lesssim \lambda_{\rm s}$ in Sec.~\ref{sec:disc}. For now, we improve on the above treatment by including the effects of anisotropy and the presence of a guide field $\boldsymbol{B_0}$.

\subsection{Effects of eddy anisotropy}\label{sec:aniso}
On scales where the anisotropy of the eddies becomes pronounced, the parallel elongation scale ($l_\parallel$) of the structure must be distinguished from its perpendicular ($l_\perp$) counterpart. As a reference, we assume a Goldreich-Sridhar scaling $l_\parallel\simeq \ell_{\rm c}^{1/3}l_\perp^{2/3}$~\cite{95GS}. For the time being, we maintain our assumption of equal parallel and perpendicular scattering mean free paths, therefore considering large-amplitude turbulence $\delta B/B_0\gtrsim1$. In this limit, the cascade operates in a regime of critical balance starting from the outer scale.

On scales $l_\parallel < \lambda_\parallel$, but $l_\perp > r_{\rm g}$, particles explore the velocity structures while gyrating around the local mean field, and hence Eqs.~(\ref{eq:Ap-interm}) and (\ref{eq:Dp-interm}) apply. The decorrelation rate becomes the maximum of $\Delta v_\parallel/l_\parallel$ and $\delta u_{l_\perp}/l_\perp$, with $\Delta v_\parallel$ the relative parallel velocity between the particle and the structure. The ratio of these two terms reads $(\Delta v_\parallel/\delta u_{l_\perp})(l_\perp/l_\parallel)$. In the regime of strong anisotropy, the ratio $l_\perp/l_\parallel$ can take small values and make the above ratio less than unity. In Alfvénic turbulence, critical balance notably implies $\delta u_{l_\perp} l_\parallel/l_\perp \simeq v_{\rm A}$, with $v_{\rm A}$ the Alfvén velocity, meaning that the ratio becomes $\Delta v_\parallel/v_{\rm A}$. Depending on $v$, the pitch-angle cosine $\mu$ and the parallel velocity of the structure, this ratio can take values below or above unity. The consequences in each limiting case are as follows.

If $\nu_{\rm f}\sim v/l_\parallel$, then the results obtained in Sec.~\ref{sec:interm} apply, especially Eq.~(\ref{eq:D-interm-2}). If, however, $\nu_{\rm f}\sim \delta u_{l_\perp}/l_\perp$, then Eq.~(\ref{eq:D-interm-2}) is modified as
\begin{align}
    \frac{A_{p\vert\,l}}{p}\,\sim\,\frac{D_{pp\vert\,l}}{p^2}&\,\sim\, \left\{ \frac{c^2}{v^2}\langle {a_\parallel}_l^2\rangle + \langle {\Theta_\parallel}_l^2\rangle + \langle {\Theta_\perp}_l^2\rangle\right\}\nonumber\\
    &\quad\quad\,\times R_{\omega/\nu}\,\frac{l_\perp}{\delta u_{l_\perp}}\,.
    \quad\quad (l_\parallel<\lambda_\parallel)
    \label{eq:D-interm-2-aniso}
\end{align}
with $R_{\omega/\nu}\simeq 1$ in that limit. It is expected that $\Theta_\perp$ will dominate in that limit, as it involves sharper perpendicular gradients.

On larger scales, $l_\parallel > \lambda_\parallel$ (with $\lambda_\parallel \sim \lambda_{\rm s}$ in large-amplitude turbulence), the value of $\nu_{\rm f}$ is determined by the competition between $v^2/(\nu_{\rm s}l_\perp^2)$ and $\delta u_{l_\perp}/l_\perp$, thus as before up to the replacement $l\rightarrow l_\perp$. The scaling with $l_\perp$ suggests that the former term ($v^2/\nu_{\rm s}l_\perp^2$) dominates on small scales, the latter ($\delta u_{l_\perp}/l_\perp$) on large scales. The transition scale can be written as before $l_{\kappa_\perp}$, such that $\kappa_{\rm s} = l_{\kappa_\perp} \delta u_{\kappa_\perp}$. The results obtained in Sec.~\ref{sec:large} then remain unchanged, provided one makes the substitutions $l\rightarrow l_\perp$ in Eq.~(\ref{eq:AD-large-2}).

As previously mentioned, the scale $l_{\kappa_\perp}$ can be written in the form $l_{\kappa_\perp}\simeq(\kappa_{\rm s}/\ell_{\rm c}\delta v_{\rm c})^{3/4}\,\ell_{\rm c}$ (assuming $q=5/3$). The condition $l_{\kappa_\perp} < \ell_{\rm c}$ thus restricts the above regime to particles of sufficiently low mean free path, i.e. sufficiently low rigidity.

\subsection{Strong guide field, $\delta B/B_0<1$}\label{sec:gfield}
The presence of a strong guide field, $\delta B/B_0 < 1$, introduces several key differences compared to the previous case. First, the cascade is generically weak on the outer scale, then it transitions to the critically balanced regime at a length scale $l_{\rm CB} \simeq \ell_{\rm c} {\mathrm M}_A^2$~\cite{2020PhPl...27a2305L}, where ${\mathrm M}_A\equiv \delta v_{\rm c}/v_{\rm A} \simeq \delta B/B_0$ defines the turbulent Alfvénic Mach number. On (perpendicular) length scales $l_\perp > l_{\rm CB}$, modes exhibit a parallel length scale $l_\parallel \simeq \ell_{\rm c}$ and $\delta u_\perp \simeq (l_\perp/\ell_{\rm c})^{1/2}\delta v_{\rm c}$. On shorter length scales, $l_\perp < l_{\rm CB}$, critical balance implies $l_\parallel \simeq \ell_{\rm c}^{1/3}l_\perp^{2/3}{\mathrm M}_A^{-4/3}$ and $\delta u_{l_\perp} \simeq \delta v_{\rm c} (l_\perp/\ell_{\rm c})^{1/3} {\mathrm M}_A^{1/3}$. At ${\mathrm M}_A<1$, the anisotropy is thus enhanced. Second, it becomes essential to distinguish the parallel diffusion coefficient ($\kappa_\parallel$) from the perpendicular one ($\kappa_\perp$). Field line diffusion relates these two coefficients as, approximately, $\kappa_\perp \sim {\mathrm M}_A^4\,\kappa_\parallel$~\cite{1993A&A...279..278C,2002PhRvD..65b3002C,2022ApJ...926...94M}. We alternatively employ effective mean free paths in the parallel ($\lambda_\parallel \equiv \kappa_\parallel/v$) and perpendicular ($\lambda_\perp \equiv \kappa_\perp/v$) directions. In this configuration, it becomes natural again to orient $e_1$ along the guide field $\boldsymbol{B_0}$. One must however distinguish the various force terms $a_1$, $\Theta_1$, $a_2$, $\Theta_2$, $\sigma_{12}$, $\sigma_{23}$ and permutations $2\leftrightarrow 3$. The shear $\sigma_{21}$ involves weaker longitudinal gradients than its counterpart $\sigma_{12}$ (recalling that $\sigma_{ij} \propto \partial_j {v_E}_i$), and it can therefore be omitted. 

The diversity of length scales prevents an exhaustive discussion of all possibilities. We thus focus on the main features and the dominant contributions, highlighting the following key differences with the previous case $\delta B/B_0\sim 1$: (i) the parallel mean free path can take large values, larger than $\ell_{\rm c}$ for a broad range of particle rigidities, because $\kappa_\parallel \propto {\mathrm M}_A^{-2}$ is generic; (ii) the transition scale $l_{\kappa_\perp}$ at which $\kappa_\perp = l_{\kappa_\perp} \delta u_{l_{\kappa_\perp}}$ is significantly reduced and likely smaller than $\lambda_\parallel$, because the perpendicular diffusion coefficient is itself sharply reduced at ${\mathrm M_A}<1$.

As a result, the typical scenario that we consider for particles of low rigidity is $ l_\parallel < \lambda_\parallel$ (large mean free path) and $l_\perp/\delta u_{l_\perp} > l_\parallel/\Delta v_\parallel$, meaning that decorrelation occurs either by escape in the perpendicular direction, or over an eddy turn over time. As before, the dominant contribution is obtained on scale $l_{\kappa_\perp}$,
and given by Eq.~(\ref{eq:AD-large-2}), provided that $l$ is replaced by $l_{\kappa_\perp}$, and that $l_{\kappa_\perp}$ is  calculated in terms of the perpendicular spatial diffusion coefficient using the above scalings. This describes a situation in which the particle moves ballistically along the field line, yet diffuses in the transverse direction at a rate controlled by perpendicular turbulent diffusion, and over a timescale comparable to the eddy turnover time.

\section{Discussion and conclusions}\label{sec:disc}

\subsection{Summary, scalings and interpretation}\label{sec:summ}
The main objective of the present paper has been twofold: (i) to introduce an effective theory designed to model stochastic Fermi-type acceleration in a generic environment; and (ii) to apply it to calculate the mean advection $A_p\equiv \langle\Delta p\rangle/\Delta t$ and diffusion $D_{pp}\equiv \langle\Delta p^2\rangle/\Delta t$ coefficients characterizing stochastic Fermi acceleration in MHD turbulence. This scheme derives an exact microscopic equation for the evolution of the momentum of a charged particle, Eq.~(\ref{eq:dyn_p}), which relates the rate of momentum change to the spatio-temporal gradients of the velocity field $\boldsymbol{v_E}$ of the frame \Rco\ in which we track the momentum of the particle along its trajectory. This generalized Fermi picture subsumes all forms of non-resonant energization in ideal electric fields, e.g. betatron acceleration, curvature and gradient drifts, transit-time damping, magnetic pumping etc., all emerge as limiting cases of Eq.~(\ref{eq:dyn_p}). To compute the transport coefficients, this equation has been supplemented with a statistical description of particle transport through the structures and of the decorrelation of the random force (Sec.~\ref{sec:transport}), drawing from and expanding on previous studies in this field. 

The transport coefficients exhibit different behaviors and scalings as a function of scale ($l$) depending on the nature of the velocity structure. One key difference with the standard picture of stochastic acceleration is that particles gain or lose energy in a systematic way while inside a velocity structure, and this at an exponential rate, because the microscopic equation takes the form $\dot p = \Gamma_l\, p$, with $\Gamma_l$ representing the random force field characterizing the structure (i.e, a gradient of $\boldsymbol{v_E}$ on scale $l$). Particle energization in regions characterized by a shear of the velocity field, or its acceleration, is hindered by pitch-angle scattering, because the sign of the force acting on the particle changes with the direction of the momentum. Conversely, in structures describing velocity field compression, which encompass the curvature of a magnetic field line (e.g., curvature drift) or the compressive modes (including gradient drift and betatron), the rate of energy gain is governed solely by the polarity of the gradient, not by the direction of the momentum. The energy gain is then constrained by the amount of time over which the particle can remain in the structure, and is thus determined by the crossing time, the trapping time, or by the lifetime of the structure itself.

On average, interactions provide a net energy gain. This net (positive) advection in momentum space arises from the influence of the structure on the momentum and on its direction. As the momentum increases, so does the energization rate, and vice versa, which leads to a net drift (the noise-induced drift of the stochastic process). Additionally, the velocity structure acts on the momentum direction and modifies the trajectory of the particle in such a way as to reduce energy loss or increase energy gain, as the case may be.  Viewed in the laboratory frame, the  effect illustrates the fact that particles interacting with random electric fields $\boldsymbol{\delta E}$ are deflected toward the direction $q\boldsymbol{\delta E}$ that promotes energy gain, thus ensuring a net average energy gain of order $\delta E^2$, as the diffusion coefficient itself. Both effects are second order $O(\Gamma_l^2)$ in the amplitude of the structure. This implies that $A_p\sim D_{pp}/p$, i.e. the system approximately satisfies detailed balance -- meaning $A_p = \left(2D_{pp} + p\,{\rm d}D_{pp}/{\rm d}p\right)/p$ -- but not exactly.

This formalism is applied to MHD turbulence in Sec.~\ref{sec:MHDturb}. The contributions of different types of velocity gradients (structures) to the diffusion coefficient are evaluated scale by scale to identify the dominant ones. Different scalings emerge in relation to the characteristic eddy velocity on the outer scale ($\delta v_{\rm c}$) and to particle rigidity (ratio of its gyroradius $r_{\rm g}$ to the coherence scale $\ell_{\rm c}$), see Fig.~\ref{fig:f1} for an illustration. This reveals a rich phenomenological landscape of applications. We now review these contributions, and then discuss their relative importance in different turbulent regimes. 

In the intermediate range of scales, $r_{\rm g}\lesssim  l\lesssim \lambda_\parallel$, i.e. above the particle gyroradius but below the parallel scattering mean free path, particles gyrate around the local magnetic field while exploring structures characterized by field lines of evolving curvature ($\Theta_\parallel$), perpendicular velocity gradients ($\Theta_\perp$), and the acceleration ($a_\parallel$) of the velocity field $\boldsymbol{v_E}$; the latter is typically subdominant. The 
dominant contribution to the diffusion coefficient, whose expression is given by Eq.~(\ref{eq:D-interm-2}) comes from the smallest scales $l\sim r_{\rm g}$. Assuming $q=5/3$, i.e. $\delta u_l \propto l^{1/3}$, and gradients scaling as $\delta u_l/l$, this results in a generic scaling $D_{pp} \sim p^2\,\delta v_{\rm c}^2/(v\ell_{\rm c}) \, (r_{\rm g}/\ell_{\rm c})^{-1/3}$. We emphasize that this calculation of the diffusion coefficient assumes it is well defined, which may not always hold. In Refs.~\cite{2021PhRvD.104f3020L,2022PhRvL.129u5101L}, in particular, we have shown that the pdf of momentum jumps in this range of scales displays hard power-law tails in large-amplitude turbulence $\delta B/B_0\gtrsim 1$. Consequently, energy gains per interaction larger than unity are possible, rendering $D_{pp}$ ill-defined. This appears limited to large-amplitude turbulence (as discussed in a forthcoming companion publication), thus the above value of $D_{pp}$ provides a reasonable approximation for that range of scales whenever $\delta B/B_0 \lesssim1$. 

In the range of scales $l > \lambda_{\rm s}$ ($\lambda_{\rm s}$ representing the particle mean free path for isotropic scattering, assuming $\delta B/B_0\gtrsim 1$), particles now explore the structures in a diffusive manner. The nature of acceleration shifts and takes the form of magnetic pumping in compressive modes, of turbulent shear acceleration, or scattering in an accelerating/decelerating velocity field, which remains subdominant due to its higher-order dependence on the turbulent velocity. As noted before, the interaction with sheared velocity gradients is constrained by the scattering of particles. A significant contribution to turbulent shear acceleration arises from the lower end of the range of scales, $l\sim \lambda_{\rm s}$, with $D_{pp} \sim p^2 \langle \sigma_l^2\rangle \lambda_{\rm s}/v$, Eq.~(\ref{eq:AD-large-1}), leading to a general scaling $D_{pp} \sim p^2 \delta v_{\rm c}^2/(\ell_{\rm c}v)\,(\lambda_{\rm s}/\ell_{\rm c})^{-1/3}$. Given that $\lambda_{\rm s}$ depends weakly on rigidity, e.g. $\lambda_{\rm s}\propto r_{\rm g}^{1/3}$, this results in $D_{pp} \sim p^2 \delta v_{\rm c}^2/(\ell_{\rm c}v)$. This result also assumes $v> \delta u_{\lambda_{\rm s}}$, where $v$ represents the particle velocity, and $\delta u_{\lambda_{\rm s}}$ denotes the characteristic velocity of eddies on the scale $\lambda_{\rm s}$.  

As the scale $l$ increases above $\lambda_{\rm s}$, the diffusive escape time of the particle becomes increasingly prolonged, eventually becoming comparable to the turn-around time of the eddy ($\sim l/\delta u_l$). This occurs at scale $l_\kappa$ , where turbulent diffusion becomes as efficient as angular scattering in promoting spatial transport, specifically $l_\kappa \delta u_{l_\kappa}\simeq \lambda_{\rm s} v$. Turbulent diffusion, with spatial diffusion coefficient $\sim l\delta u_l$, then regulates particle diffusion on scales $l\gtrsim l_\kappa$, and transports the particles through the velocity gradients for durations comparable to an eddy turn-around time. In the case of shear acceleration, this leads to $D_{pp} \sim p^2 \delta v_{\rm c}^3/(\ell_{\rm c}v^2)$, see Eq.~(\ref{eq:AD-large-2}). 

In compressive modes, the same effect provides the dominant contribution at the scale $l_\kappa$, yielding the simple result $D_{pp} \sim p^2 \delta u_{l_\kappa}/l_\kappa$ [Eq.~(\ref{eq:AD-large-2})]. The scaling differs from that of the turbulent shear contribution because the energy gain is no longer hindered by particle scattering.

These considerations remain valid even after accounting for eddy anisotropy and for the presence of a guide field, with the following modifications. First, the perpendicular eddy scale $l_\perp$ now sets the characteristic scale that controls the scalings of $D_{pp}$, i.e. $l\rightarrow l_\perp$. The presence of anisotropy opens an interesting regime where $l_\perp \gtrsim r_{\rm g}$ while the parallel scale $l_\parallel \lesssim \lambda_\parallel$ ($\lambda_\parallel$ the parallel scattering mean free path). The enhanced parallel length scale (relative to $l_\perp$) extends the confinement time, potentially up to the eddy lifetime, see Sec.~\ref{sec:aniso} for specific conditions of application. A guide field increases the parallel mean free path, possibly up to the point where particles can cross a coherence volume of the turbulence without scattering. Consequently, the regime $l_\parallel \lesssim \lambda_\parallel$ becomes prevalent. The reduced ratio of perpendicular to parallel scattering mean free path, as well as the reduced ratio of perpendicular to parallel length scales of the eddy imply that the confinement time can become limited by escape in the perpendicular direction, possibly through the influence of turbulent diffusion from smaller scale eddies. This regime notably arises whenever the relative velocity between the particle and the structure along the magnetic field is sufficiently small. The dominant contribution then emerges at scales $l_\perp \sim l_{\kappa_\perp}$, with as above -- up to the account of anisotropy in transport -- $l_{\kappa_\perp} \delta u_{l_{\kappa_\perp}}\sim \kappa_\perp$, the perpendicular diffusion coefficient. On this scale, particles drift along the mean magnetic field, collect the influence of field line curvature and perpendicular gradients, and exit the structure in the perpendicular direction over an eddy turn-around time. The resulting diffusion coefficient becomes $D_{pp}\sim p^2 \delta u_l/l$ at $l\sim l_{\kappa_\perp}$, [Eq.~(\ref{eq:D-interm-2-aniso})], similar to that of compressive large-scale modes discussed earlier.

With the exception of modes on scale $l\sim r_{\rm g}$, the above diffusion coefficients exhibit weak dependency on momentum beyond the prefactor $p^2$, indicating that the ``hard-sphere'' ($D_{pp}\propto p^2$) regime is generic. However, the contributions exhibit different scalings relative to the turbulent velocity on the outer scale, more specifically $\propto \delta v_{\rm c}^m$ with $m=1$, $2$, or $3$ for the dominant contributions. For $\delta B/B_0 \gtrsim 1$, the generic scaling is $m=2$ in small-scale modes ($l\lesssim \lambda_\parallel$), but $m=2$ or $3$ for shear modes on large scales. For large-scale compressive modes in strong turbulence ($\delta B/B_0\gtrsim 1$), or in regions with curved field lines and perpendicular gradients in $\delta B/B_0<1$ turbulence with significant eddy anisotropy -- i.e.,  situations where particles can be confined within the structure for an entire eddy turn-around time -- the dominant contribution scales as $m=1$. This finding is particularly significant in sub-relativistic turbulence, as it indicates faster acceleration than expected from the naive scaling $D_{pp}\propto (\delta v_{\rm c}/c)^2$. This fast acceleration regime, however, assumes $l_\kappa \leq \ell_{\rm c}$ (or $l_{\kappa_\perp} \leq \ell_{\rm c}$), which limits it to the lower range of particle rigidities, such that 
\begin{equation}
    \frac{l_\kappa}{\ell_{\rm c}}\,\simeq\,\left(\frac{\lambda_{\rm s}}{\ell_{\rm c}}\right)^{3/4}\left(\frac{v}{\delta v_{\rm c}}\right)^{3/4}\,\leq \, 1\,.
    \label{eq:lk}
\end{equation} 
For particles with rigidities sufficiently large to violate this bound, the contribution to the diffusion coefficient reduces to $D_{pp}\sim p^2 \delta v_{\rm c}^2/(\lambda_{\rm s}v)$. The transition between these two regimes is smooth, because $l_\kappa=\ell_{\rm c}$ means $\lambda_{\rm s}v = \ell_{\rm c} \delta v_{\rm c}$.

Clearly, the prevailing regime also depends on the contribution of different modes to the turbulence. The above discussion suggests that compressive modes, if present, are likely to dominate. Linear Alfvén waves induce shear and acceleration/deceleration of field lines to linear order, but they do not contribute to $\Theta_\parallel$, $\Theta_\perp$ or $\Theta$~\cite{2021PhRvD.104f3020L}. However, they can seed compressive modes to second order, which could therefore contribute with a scaling up to $m=3$ (if particles are trapped in these modes).
Distinguishing the contributions and determining the various numerical prefactors in the above estimates requires dedicated measurements of the statistics of the various force terms,  scanning different turbulence parameters (e.g. Alfvénic and sonic Mach numbers). Moreover, in high-beta collisionless plasmas, pressure anisotropies may feedback on plasma motions through magneto-immutability~\cite{2019JPlPh..85a9014S,2023PhRvX..13b1014A}, thereby directly influencing the scaling of structures responsible for acceleration. This underscores the need for dedicated numerical studies of this regime of turbulence.

If the diffusion coefficient scales as $D_{pp}\propto (\delta v_{\rm c})^m$, the characteristic acceleration timescale scales as $\tau_{\rm acc}\equiv p^2/D_{pp} \propto \delta v_{\rm c}^{-m}$. Meanwhile, the turbulent cascade leads to plasma heating through dissipative processes on a timescale $\tau_{\rm heat} \sim \beta \ell_{\rm c}/v_{\rm A}$. Thus, in regions where $m\simeq1$, Fermi acceleration in velocity gradients can provide a competitive contribution to plasma heating, even in sub-relativistic turbulence.

Throughout, we have assumed that the relative energy gain per interaction is less than unity, although this may not always hold, even for well behaved pdfs of the random force terms. Consider, for example, the (compressive) result $D_{pp} = x_{\Theta}\,p^2\,\delta v_{\rm c}/\ell_{\rm c}$, to which we have added a prefactor $x_{\Theta}<1$; $x_\Theta$ accounts for the numerical prefactor absent in the previous estimates and for the energy content of compressive modes. The average momentum gain per interaction with a structure can be expressed as $\langle\Delta p^2\rangle^{1/2} \simeq (D_{pp}\Delta t)^{1/2}$, with $\Delta t$ being the interaction time, here $\ell_{\rm c}/\delta v_{\rm c}$ (eddy turn-around time, or eddy crossing time through turbulent diffusion). For the above scaling, $\langle\Delta p^2\rangle^{1/2}/p \simeq  x_{\Theta_l}$, so in specific regions where the velocity gradient is sufficiently large, e.g. $x_\Theta^{-1}$ above the mean, the relative jump in momentum can surpass unity. In such  cases, however, the actual momentum variation becomes even more substantial, because the energy gain is truly exponential, implying that $\Delta p/p$ should rather be interpreted as $\Delta \ln p$. Consequently, the diffusion process becomes heterogeneous, featuring pronounced momentum jumps in localized regions and moderate gains elsewhere. 

\subsection{Modeling transport}\label{sec:transport}
The Fokker-Planck formalism serves as the standard framework for tracking the evolution of the distribution function of particles undergoing stochastic acceleration. However, as exemplified in the above discussion, this formalism may not always be adequate. Although central limit convergence should ensure that the random walk behaves as Brownian motion on asymptotic timescales, as illustrated in Appendix~B of \cite{2021PhRvD.104f3020L}, the timescale for this transition must be considered. If this timescale is comparable to or shorter than the other characteristic timescales of the system, the Fokker-Planck formalism can be employed using the advection and diffusion coefficients calculated above. Otherwise, alternative formulations must be sought to track the distribution function over time.

In the case of heterogeneous diffusion discussed above, one can employ the formalism of Refs.~\cite{2006ApJ...642..244M,2020MNRAS.499.4972L}, which models the plasma as composed of two subpopulations, one (labeled 1) that undergoes acceleration, and another (labeled 0) that does not. These subpopulations exchange particles at rates $\nu_{0\rightarrow 1}$ and $\nu_{1\rightarrow 0}$. The population $0$ is termed ``trapped'' in Ref.~\cite{2020MNRAS.499.4972L} because it must wait for a time $\sim \nu_{0\rightarrow 1}^{-1}$ before entering a region where acceleration is effective. The connection to the present findings is straightforward: particles $1$ are those that populate regions where $\Delta p/p>1$ (leading to exponential energy gains), and are converted into particles of population $0$ at rate $\nu_{\rm f}$ ($\nu_{\rm f}$ represents the decorrelation rate in Sec.~\ref{sec:MHDturb}). The rate $\nu_{0\rightarrow 1}$ is determined by the filling fraction of regions $1$. Reference~\cite{2020MNRAS.499.4972L} then solves a system of two equations, one for population $0$ with entry ($\nu_{1\rightarrow 0}$) and exit ($\nu_{0\rightarrow 1}$) rates and no energizing term, and  another for population $1$ with reversed entry and exit rates and a Fokker-Planck term to model stochastic acceleration. This formalism enables the description of heterogeneous diffusion in a binary mixture and could be generalized to include more acceleration rates (i.e., additional mixtures or populations), albeit at the cost of additional parameters. For momentum-independent entry, exit and acceleration rates, Ref.~\cite{2020MNRAS.499.4972L} demonstrates that the distribution function typically exhibits a power-law form with an index determined by the ratios of these rates, reminiscent of the Fermi formula determining the spectral index as a function of the ratio of acceleration and escape timescales~\cite{1954ApJ...119....1F}. This problem has also been analytically investigated in Ref.~\cite{2021PhRvD.104f3020L}, which models the random walk as a succession of energy gains or losses in highly localized regions and derives the time-dependent distribution function using a large-deviation argument. Notably, it illustrates how the power-law tail gradually transitions to a Gaussian distribution on asymptotic timescales, once all particles have experienced at least one acceleration site (i.e., on a timescale $\gg \nu_{0\rightarrow 1}^{-1}$). Detailed measurements of these rates in numerical simulations would provide valuable insights to progress along these lines. 

The Fokker-Planck approximation also breaks down whenever the diffusion or advection coefficients become ill-defined, for instance, when the pdf of the random force $\Gamma_l$ displays hard tails. As discussed in previous works~\cite{2021PhRvD.104f3020L,2022PhRvL.129u5101L}, this situation arises in large amplitude turbulence ($\delta B/B_0 \gtrsim 1$) for the terms $\Theta_\parallel$ and $\Theta_\perp$ that govern particle acceleration in the range of scales $l\lesssim \lambda_{\rm s}$. These hard tails are linked to the peculiar statistics of the curvature scalar $\kappa \equiv \vert \boldsymbol{b}\cdot\boldsymbol{\nabla}\boldsymbol{b}\vert$ of magnetic field lines, whose pdf displays a power-law tail $\propto \kappa^{-2.5}$ at values of $\kappa$ well above the mean. This behavior has been measured in MHD simulations~\cite{2001PhRvE..65a6305S,2019PhPl...26g2306Y,2020ApJ...898...66Y} and in the solar wind magnetosheath~\cite{2020ApJ...893L..25B,2020ApJ...898L..18H}. In such a situation, the evolution of the average distribution function can be described using a master equation for Markov processes. This equation generalizes the Fokker-Planck equation by accounting for the full pdf of the momentum jumps~\cite{2017PhRvL.119d5101I,2022PhRvL.129u5101L}. 

The above discussion suggests that heterogeneous momentum space diffusion may be a common occurence in MHD turbulence. In the sub-relativistic regime, with $\delta B/B_0\lesssim 1$, this results from the scaling of the diffusion coefficient $\propto \delta v_{\rm c}/\ell_{\rm c}$ observed over a range of scales. In the limit $\delta B/B_0 \gtrsim 1$, this follows from the hard tails of the statistics of momentum jumps in curved field lines. There remains the intermediate limit of trans-relativistic and large amplitude turbulence, characterized by $\delta v_{\rm c}\sim c$ and $\delta B_0\lesssim B$, where all discussed scalings become comparable, and where the average energy gain per interaction remains consequent. 

\subsection{Comparison to numerical simulations}\label{sec:simnum}
In the sub-relativistic limit, heterogeneous diffusion has been observed in  numerical simulations of MHD turbulence with test particles at amplitudes $\delta B/B_0 \sim 0.1-0.2$~\cite{2020ApJ...894..136T}. The detailed analysis of particle trajectories in this work indicates that acceleration is rapid in certain localized regions where particles are trapped, with the energy gain attributed to curvature drift acceleration ($\Theta_\parallel$ in our notations). Particle trapping, accompanied by large energy gains $\Delta p/p > 1$, has also been observed in a MHD simulation of incompressible turbulence at $\delta B/B_0\simeq1$~\cite{2022ApJ...928...25P}. The authors demonstrate that most trapped/accelerated particles exhibit a low pitch-angle cosine, indicating that they are trapped because of their slow motion along the mean magnetic field at that location. While trapped, they undergo acceleration at an exponential rate over a time $\sim \ell_{\rm c}/\delta v_{\rm c}$ (in our notations). Our findings are consistent with these results. In Ref.~\cite{2023ApJ...959...28P}, the authors perform several compressible MHD simulations with and without nonlinear effects, and with and without phase randomization, at a characteristic amplitude $\delta B/B_0 \simeq 0.1$. They conclude that structures (as opposed to random-phased linear waves) contribute significantly to particle acceleration. These authors also report super-diffusive behavior in momentum space and exponential acceleration for some particles, which they interpret as the contribution of compression perpendicular to the magnetic field. They interpret the maximal energy gain as limited by escape out of the structure in the perpendicular direction, which they associate with the effect of fluctuations. This general result agrees, at least qualitatively, with our present findings in the presence of a guide field, see Sec.~\ref{sec:gfield}.

Particle acceleration has been extensively studied in kinetic simulations of relativistic turbulence~\cite{17Zhdankin,2018ApJ...867L..18Z,18Comisso,2019ApJ...886..122C,2019PhRvL.122e5101Z,2020ApJ...893L...7W,2021ApJ...922..172Z,2022ApJ...924L..19V,2022PhRvD.106b3028B,2025arXiv250203042W}. The momentum distribution functions observed in these simulations display power-law spectra, with spectral indices that become harder as $u_{\rm A}/c$ and $\delta B/B_0$ increase. By directly measuring the advection and the diffusion coefficients through particle tracking in their simulations, the authors of Refs.~\cite{2020ApJ...893L...7W,2025arXiv250203042W} have been able to reproduce the observed particle distributions using a Fokker-Planck approach. An interesting observation is that these advection and diffusion coefficients do not observe detailed balance. 

The present calculations can be extended to the relativistic regime with the following rules: (i) the diffusion process must be tracked at all times in the frame \Rco, which moves relative to the laboratory frame at characteristic (relativistic) velocity $\sim v_{\rm A}$, and the diffusion coefficient is indeed measured in this way in the references cited above; (ii) the velocity gradients involve the four-velocity ${u_E}^\mu$, implying that all $\langle \Theta^2\rangle$ etc. scale as $u_{\rm A}^2$; (iii) turbulent diffusion, however, involves the three-velocity, just as any eddy turn-around time; (iv) the ratio $v/\delta u_l$ is no longer generically larger than unity, which implies that particle trapping in structures can become more frequent than in the sub-relativistic limit where $\delta u_l\ll c$. Consequently, the diffusion coefficients derived in Sec.~\ref{sec:MHDturb} exhibit generic scalings $D_{pp} \sim (u_{\rm A}/c)^2 p^2/\ell_{\rm c}$, consistent with those measured in numerical simulations of relativistic turbulence. One possible exception is the small-scale regime $l\sim r_{\rm g}$, which provides the dominant contribution in the absence of trapping inside structures, as discussed in Sec.~\ref{sec:aniso}. This regime yields an additional scaling $\propto (r_{\rm g}/\ell_{\rm c})^{-1/3}$, although this scaling is likely modified by the influence of trapping and the heavily tailed pdf of field line curvature. Furthermore, a detailed comparison between the observed energy histories (in the comoving frame \Rco) and those predicted by the present generalized Fermi process, specifically Eq.~(\ref{eq:pp-interm}) describing gyration around local magnetic field lines, demonstrates that the model captured by this equation can reproduce most of the observed energy gain for low-energy particles~\cite{2022PhRvD.106b3028B}. Lastly, particle acceleration due to interactions with 2D velocity gradients in the plane perpendicular to the guide field $\boldsymbol{B_0}$ explains why acceleration appears equally efficient in 2D as in 3D in the aforementioned numerical simulations, while gyroresonant wave-particle interactions are suppressed in a strict 2D setting with an out-of-plane guide field.

\subsection{Conclusions}
Existing numerical simulations indicate that particle acceleration does not proceed smoothly or uniformly, even in small-amplitude, sub-relativistic turbulence, and that most of the energization appears associated with localized regions of strong gradients. As discussed in the Introduction, distinguishing between such structures and wavepackets can be ambiguous. For instance, a large-amplitude slow mode wavepacket with an extent $l$ not exceeding several wavelengths behaves as a structure and generates perpendicular compressions that can significantly influence particle acceleration, as suggested by recent numerical simulations. Similarly, a large-amplitude, short-extent perturbation with a polarity of the form $\boldsymbol{\delta v}/v_{\rm A}\simeq \pm \boldsymbol{\delta B}/B_0$ would be categorized as Alfvénic. However, due to its limited extent, it would not induce resonant wave-particle interaction. The critical question is how to model the impact of such modes or structures on particle transport in momentum space. The present formalism has the ambition to provide a way forward by capturing all forms of non-resonant acceleration. The central equation, Eq.~(\ref{eq:dyn_p}), is exact within the framework of ideal MHD. In several ways, the procedure that we follow to calculate the transport coefficients is analogous to that used in extended quasilinear calculations, except that the standard equation connecting $\dot p$ to $\boldsymbol{\delta E}$ is replaced here by Eq.~(\ref{eq:dyn_p}). Although the discussion has primarily focused on the application of this formalism to MHD turbulence, it can be used to model acceleration in more complex geometries, such as a turbulence composed of flux ropes, multi-island reconnecting regions, and other forms of turbulence initiated by various instabilities. Therefore, this formalism is applicable to a wide range of environments. 

Based on the calculations presented in Sec.~\ref{sec:MHDturb} and existing numerical studies of particle acceleration in MHD turbulence, one is tempted to outline a global picture as follows. In large amplitude turbulence, $\delta B/B_0\gtrsim 1$, the energization process appears dominated by the heavy-tailed pdfs of momentum jumps associated with particle acceleration in strongly curved and pinched field lines, as argued earlier~\cite{2022PhRvL.129u5101L}. For $\delta B/B_0\lesssim 1$ and/or sub-relativistic turbulence, compressive modes play a crucial role. These modes can be perpendicular to the magnetic field or represent an isotropic average (if $\delta B/B_0\sim 1$). Acceleration becomes efficient when particles are confined in regions of strong velocity gradient over an eddy turn-around time, in which case they escape through turbulent diffusion induced by smaller scale eddies. The diffusion coefficient can then scale linearly with the characteristic eddy velocity, allowing particles to gain substantial amounts of energy ($\Delta p/p>1$) in localized regions. This results in rapid acceleration and efficient plasma heating in regions of strong gradients in the velocity of magnetic field lines. In the absence of compressive modes, acceleration  occurs primarily in the turbulent shear component of the velocity flow. While this overall picture aligns qualitatively with recent numerical simulations, further investigation is warranted, particularly to explore the regime of small rigidities $r_{\rm g}/\ell_{\rm c}$, where the effects of turbulence anisotropy become more pronounced.

\begin{acknowledgments}
It is a pleasure to thank L. Vlahos, D. Uzdensky and V. Zhdankin for insightful discussions, as well as the participants of the 14th Plasma Kinetics Working Meeting organized by A. Schekochihin and M. Kunz at the Wofgang Pauli Institute in Vienna. The author thanks both anonymous referees whose suggestions have helped to improve and clarify the manuscript.
\end{acknowledgments}

\appendix

\bibliography{refs}

\end{document}